\newif\iffigs\figstrue

\documentstyle[12pt]{article}
\iffigs
\else
\fi

\batchmode
\newfont{\footscrfont}{rsfs10}
  \newfont{\footbbbfont}{msbm10}
  \newfont{\manfont}{manfnt}
\errorstopmode

\newif\ifscrf\scrftrue
\ifx\footscrfont\nullfont
  \scrffalse
\fi

\newif\ifamsf\amsftrue
\ifx\footbbbfont\nullfont
  \amsffalse
\fi

\def\ppnumber{\vbox{\baselineskip14pt\hbox{hep-th/9807040}
\hbox{CU-TP-899}
\hbox{CLNS-98/1566}}}
\def\ppdate{July 1998}
\def\pplogo{\vbox{\kern-\headheight\kern -15pt
\halign{##&##\hfil\cr&{
\ppnumber}\cr\rule{0pt}{2.5ex}&\ppdate\cr}
}}

\makeatletter
\date{}
\def\dedicatory#1{\def\@date{\normalsize\it#1}}
\def\subjclass#1{\def\@thefnmark{}\@footnotetext{1991
    {\it Mathematics Subject Classification.} #1}}
\def\keywords#1{\def\@thefnmark{}\@footnotetext{
    {\it Key words and phrases.} #1}}

\def\ps@firstpage{\ps@empty \def\@oddhead{\hss\pplogo}%
  \let\@evenhead\@oddhead 
}
\def\maketitle{\par
 \begingroup
 \def\thefootnote{\fnsymbol{footnote}}
 \def\@makefnmark{\hbox
 to 0pt{$^{\@thefnmark}$\hss}}
 \if@twocolumn
 \twocolumn[\@maketitle]
 \else \newpage
 \global\@topnum\z@ \@maketitle \fi\thispagestyle{firstpage}\@thanks
 \endgroup
 \setcounter{footnote}{0}
 \let\maketitle\relax
 \let\@maketitle\relax
 \gdef\@thanks{}\gdef\@author{}\gdef\@title{}\let\thanks\relax}

\def\abstract{\if@twocolumn
\section*{Abstract}
\else \small
\begin{center}
{\bf ABSTRACT}
\end{center}
\quotation
\fi}

\def\thebibliography#1{\section*{References\@mkboth
 {REFERENCES}{REFERENCES}}\small\list
 {[\arabic{enumi}]}{\settowidth\labelwidth{[#1]}\leftmargin\labelwidth
 \advance\leftmargin\labelsep
 \usecounter{enumi}}
 \def\newblock{\hskip .11em plus .33em minus .07em}
 \sloppy\clubpenalty4000\widowpenalty4000
 \sfcode`\.=1000\relax}

\newif\iffn\fnfalse

\@ifundefined{reset@font}{\let\reset@font\empty}{} 
\long\def\@footnotetext#1{\insert\footins{\reset@font\footnotesize
    \interlinepenalty\interfootnotelinepenalty
    \splittopskip\footnotesep
    \splitmaxdepth \dp\strutbox \floatingpenalty \@MM
    \hsize\columnwidth \@parboxrestore
   \edef\@currentlabel{\csname p@footnote\endcsname\@thefnmark}\@makefntext
    {\rule{\z@}{\footnotesep}\ignorespaces
      \fntrue#1\fnfalse\strut}}}

\makeatother

\ifamsf
  \newfont{\bigbbbfont}{msbm10 scaled\magstep2}
  \newfont{\bbbfont}{msbm10 scaled\magstep1}  
  \newfont{\smallbbbfont}{msbm8}
  \newfont{\tinybbbfont}{msbm6}
  \newfont{\smallfootbbbfont}{msbm7}
  \newfont{\tinyfootbbbfont}{msbm5}
  \newfont{\biggthfont}{eufm10 scaled\magstep2}
  \newfont{\gthfont}{eufm10 scaled\magstep1}  
  \newfont{\smallgthfont}{eufm8}
  \newfont{\tinygthfont}{eufm6}
  \newfont{\footgthfont}{eufm10}
  \newfont{\smallfootgthfont}{eufm7}
  \newfont{\tinyfootgthfont}{eufm5}
\fi

\ifscrf
  \newfont{\scrfont}{rsfs10 scaled\magstep1}  
  \newfont{\smallscrfont}{rsfs7}
  \newfont{\tinyscrfont}{rsfs7}
  \newfont{\smallfootscrfont}{rsfs7}
  \newfont{\tinyfootscrfont}{rsfs7}
\fi

\ifamsf
  \newcommand{\Bbb}[1]{\iffn
      \mathchoice{\mbox{\footbbbfont #1}}{\mbox{\footbbbfont #1}}
      {\mbox{\smallfootbbbfont #1}}{\mbox{\tinyfootbbbfont #1}}\else
      \mathchoice{\mbox{\bbbfont #1}}{\mbox{\bbbfont #1}}
      {\mbox{\smallbbbfont #1}}{\mbox{\tinybbbfont #1}}\fi}
  
\else
  \def\bigbbbfont{\bf}
  \def\Bbb{\bf}
  
\fi

\ifscrf
  \newcommand{\Scr}[1]{\iffn
    \mathchoice{\mbox{\footscrfont #1}}{\mbox{\footscrfont #1}}
    {\mbox{\smallfootscrfont #1}}{\mbox{\tinyfootscrfont #1}}\else
    \mathchoice{\mbox{\scrfont #1}}{\mbox{\scrfont #1}}
    {\mbox{\smallscrfont #1}}{\mbox{\tinyscrfont #1}}\fi}
\else
  \def\Scr{\cal}
\fi

\def\C{{\Bbb C}}

\def\R{{\Bbb R}}
\def\Z{{\Bbb Z}}

\def\bearray{\begin{eqnarray}}
\def\eearray{\end{eqnarray}}
\def\bearraynn{\begin{eqnarray*}}
\def\eearraynn{\end{eqnarray*}}
\def\bfig{\begin{figure}}
\def\efig{\end{figure}}

\def\opeq#1{\advance\lineskip#1 \advance\baselineskip#1
	\advance\lineskiplimit#1}

\def\cM{{\Scr M}}

\def\cD{{\Scr D}}

\def\cMc{{\hfuzz=100cm\hbox to 0pt{$\;\overline{\phantom{X}}$}\cM}}
\def\barcD{{\hfuzz=100cm\hbox to 0pt{$\;\overline{\phantom{X}}$}\cD}}

\ifamsf

\else

\fi

\def\Lie{{\rm Lie}}

\newtheorem{Proposition}{Proposition}[section]

\newtheorem{Theorem}{Theorem}[section]
\newtheorem{Lemma}{Lemma}[section]
\newtheorem{Corrolary}{Corrolary}[section]

\newcommand{\be}{\begin{equation}}
\newcommand{\ee}{\end{equation}}
\newcommand{\bea}{\begin{eqnarray}}
\newcommand{\eea}{\end{eqnarray}}
\newcommand{\nn}{\nonumber}
\newcommand{\bdm}{\begin{displaymath}}
\newcommand{\edm}{\end{displaymath}}

\newcommand{\bp}{\begin{Proposition}}
\newcommand{\ep}{\end{Proposition}}
\newcommand{\bt}{\begin{Theorem}}
\newcommand{\et}{\end{Theorem}}
\newcommand{\bl}{\begin{Lemma}}
\newcommand{\el}{\end{Lemma}}
\newcommand{\bc}{\begin{Corrolary}}
\newcommand{\ec}{\end{Corrolary}}

\begin{document}

\title{D-particles on $T^4/{\bf Z}_n$ orbifolds and their resolutions}

\author{B.~R.~Greene$^{1,a}$,\\
C.~I.~Lazaroiu$^{1,b}$ and Piljin~Yi$^{2,c}$ }

\maketitle
\vbox{
\centerline{$^1$Departments of Physics and Mathematics}
\centerline{Columbia University}
\centerline{N.Y., N.Y. 10027}
\medskip
\centerline{$^2$Newman Laboratory of Nuclear Studies}
\centerline{Cornell University}
\centerline{Ithaca, N.Y. 14850}
\medskip
\bigskip
}

\abstract{We formulate the effective field theory of a D-particle on 
orbifolds of $T^4$ by a cyclic group as a gauge theory in a $V$-bundle 
over the 
dual orbifold. We argue that this theory admits Fayet-Iliopoulos 
terms analogous to those present in the case of noncompact orbifolds. 
In the $n=2$ case, we present some evidence that turning on such terms 
resolves the orbifold 
singularities and may lead to a $K3$ surface 
realized as a blow up of the fixed points of the cyclic group action. }

\vskip .6in

$^a$ greene@phys.columbia.edu

$^b$ lazaroiu@phys.columbia.edu

$^c$ piljin@mail.lns.cornell.edu

\pagebreak

\section*{Introduction}

	The investigation of the effective short-distance geometric 
description of space-time is one of the most fascinating ways of exploring 
the fundamental properties of string theory. 
The first order perturbative analysis at weak string coupling 
(`stringy geometry') 
reduces to questions about the moduli space of (a certain class of) 
conformal field theories. This analysis reveals 
\cite{stringy_geometry} a    
rich phase structure of the conformal field theory  moduli space, 
unifying a large class of what were once thought to be unrelated models and 
allowing for topology changing transitions among them. 

	The corresponding nonperturbative analysis is still in its infancy, 
but it has been known for some time 
\cite{drastic_top_change,Witten_phases,Shenker,DKPS} that qualitatively 
new phenomena, involving rather drastic departures from the geometry 
of the classical moduli space should occur in this regime. Lacking a 
general nonperturbative formulation of string theory compactifications, 
the best one can do at this 
stage is to approach the problem by using nonperturbative objects as probes 
of the relevant moduli space, much as one  
uses conformal field theory to probe the pertubative 
approximation. In this light, 
a first understanding of how the perturbative picture of 
\cite{stringy_geometry} is corrected by nonperturbative effects can be gained 
by considering `brany-geometry', i.e. 
the moduli space of various D-branes present in a string-theoretic 
compactification. To carry out such an investigation one needs a description 
of the low energy dynamics of D-branes which is reasonably easy to manipulate. 
In the case of flat space-time, such an  effective low energy description 
is provided \cite{Witten} by the reduction of the  D=10, N=1 supersymmetric 
Yang Mills theory to the D-brane worldvolume.  
Unfortunately, the appropriate description in the case of a nontrivial 
spacetime geometry is far from being understood properly, despite various 
attempts \cite{curved}. Although some tantalizing  conjectures 
have been made by exploiting rather abstract features expected of such a 
description \cite{HM,HO,BSV}, the rigorous physical formulation of the 
problem is lacking. 

One of the processes of `brany-geometry' that one would like to 
investigate is the resolution of orbifold singularities by D-branes. 
The case of flat {\em noncompact} cyclic orbifolds of the type $C^d/\Z_n$ 
was considered in \cite{DM,JM,DGM,G,DG,Muto,Mohri}. 
There it was shown that the basic physical mechanism for the orbifold 
resolution was the posssibility to turn on certain vacuum expectation 
values in the twisted sectors of the orbifold, thus inducing nontrivial 
Fayet-Iliopoulos terms in the effective supersymmetric 
theory of the D-branes.

The present work is an attempt to understand the analogous process in the 
case of {\em compact} orbifolds.  
As we will see, the resolution process is considerably more subtle. 
The reason for this is that the compactness of the original orbifold 
leads us to a problem of moduli spaces of a certain class of gauge 
connections because the effective theory is inherently a gauge
{\it field} theory. The mathematical difficulty is comparable with the one 
encountered when generalizing from linear algebra to functional analysis. 
Namely, we will discover that both the geometric and the analytic aspects 
of our problem are nontrivial, unlike the case of \cite{DM,DGM}.

To avoid unnecessary technicalities, we will focus on the case of one 
D-particle on abelian orbifolds of $T^4$. This admits an effective 
low-energy description via a supersymmetric theory along the 
lines of \cite{Taylor}. In section 1, we review the resolution process 
of noncompact orbifolds from a point of view which is relevant for this 
paper and we explain the origin of the difficulty in the compact case.
In section 2, we formulate the relevant low-energy 
theory starting from the abstract set-up of \cite{CDS} and we show that it 
leads to a problem of equivariant gauge connections on a $V$-bundle
\footnote{The concept of a $V$ -bundle turns out to be the proper 
mathematical framework for formulating our orbifold theory. 
Although this paper can be read without any understanding of this 
concept, we will occasionally state some results in this language, since it 
allows us to give a precise meaning to the so-called `singular bundles' 
sometimes mentioned in the matrix-theory literature.  
A $V$-bundle $E$, say, over 
$T^4/\Z_n$, is simply a vector bundle $F$ over $T^4$ together with linear 
identifications $\phi_{\alpha}$ among its fibers, 
giving an action of $\Z_n$ by 
bundle automorphisms which cover the action of $\Z_n$ on $T^4$. 
Thus, for any element $\alpha$ of $\Z_n$, and for any point $x \in T^4$, one 
has a linear isomorphism $\phi_{\alpha,x}$ between the fiber $F_x$ of $F$ 
over $x$ and the fiber $F_y$ over the image $y=\alpha \cdot x$ of $x$. 
This specifies the way in which the fibers of $F$ are to be identified under 
the action of the orbifold group. One imposes the compatibility condition 
$\phi_{\beta,\alpha\cdot x}\circ \phi_{\alpha,x}=\phi_{\beta+\alpha,x}$. 
In general, the action of $\Z_n$ on $T^4$ has a number of fixed points. 
If $x_0$ is such a point, then the maps $\phi_{\alpha,x_0}$
give a representation of $\Z_n$ in the fiber $F_{x_0}$, called the {\em 
isotropy representation} at $x_0$. 
A $V$-bundle is characterized by the topological 
invariants of the underlying vector bundle and by its isotropy 
representations at the fixed points of the orbifold action. The simplest 
example of a $V$-bundle --which will occur below-- 
is the usual twisted product 
$T^4\times_{\Z_n}R$ (also called a product $V$-bundle), where $R=(\C^r,\rho)$ 
is a representation space for $\Z_n$. This has trivial underlying bundle 
$F=T^4 \times \C^r$ and $V$-structure given by the constant identifications 
$\phi_\alpha(x)=\rho(\alpha), ~\forall x \in T^4$. 
It has isotropy representations equal 
to $\rho$ at each of the fixed points. 
Given a $V$-bundle $E$, one can consider 
connections on the underlying vector bundle $F$, which are 
compatible with the $V$-bundle structure (i.e. 
with the maps $\phi_{\alpha,x}$). 
Such objects are called {\em invariant, equivariant } or $V-$ connections 
on $E$, 
and are nothing else than usual connections $A$ 
on the underlying bundle subject to the condition that their pull-backs  
$^\alpha A$ by the orbifold action are 
gauge-equivalent with $A$:
\bdm
^\alpha A=\sigma_\alpha (d+A)\sigma_\alpha^{-1}~~,
\edm
with $\sigma_{\alpha}=\phi_{\alpha, (-\alpha)\cdot x}$. 
The gauge transformations 
$\sigma_\alpha$  implementing this equivalence encode the $V$-bundle 
structure. 
If the underlying bundle $F$ is equipped with a hermitian metric such that 
$\phi_{\alpha,x}$ are unitary for all $\alpha,x$, then $E$ is called a 
hermitian $V$-bundle. In this case, those unitary automorphisms $U$ 
of $F$ which 
commute with $\phi_{\alpha}$ are called {\em V-gauge transformations} of $E$.
They satisfy the projection condition:
\bdm
U((-\alpha)\cdot x)=\sigma_\alpha(x)U(x)\sigma_\alpha(x)^{-1}~~
\edm
and act on connections $A$ in the usual manner. Then the 
projection condition satisfied by $U$ assures that the gauge transform of a 
$V$-connection is still a $V$-connection.  
For more information about $V$-bundles and invariant connections 
the reader is reffered to \cite{Vbundles}.} over the orbifold. 
In section 3, we give a clear account of how the singular moduli space 
arises in the case of $T^4/\Z_2$, which is the focus of the rest of the paper. 
In section 4, proceeding along the lines of\cite{DM,JM}, we argue that 
the orbifold gauge theory admits a certain class of loclaized 
Fayet-Iliopoulos terms, leading to a moduli space of singular 
gauge connections. In section 5  we give some evidence that this 
provides a resolution of the original singular moduli space. Finally, 
section 6 presents our conclusions and speculations.
During the preparation of this paper, we became aware of \cite{RW},  
which has some overlap with our work. 

\section{General considerations}

\subsection{The quotienting procedure}

The general method for obtaining an effective field 
theory of D-particles on a quotient space of the form 
$K=\R^4/G$, with $G$ a discrete group, is to consider a system of 
D-particles on the covering space, together with their images, and to project 
the resulting field theory onto its $G$-invariant part. The basic example 
of this procedure is the case of one D-particle over an orbifold of the type 
$\C^2/\Z_n$, which is obtained \cite{DM} by letting $G=\Z_n$ act on 
$\C^2$ via its fundamental representation $\rho_f$ and on the Chan-Paton 
factors via its 
regular representation $\rho_{reg}$. In this case, one has a system of $n$ 
images distributed over $\C^2$ in a $\Z_n$-invariant way. 
More precisely, there are two 
types of variables entering the resulting supersymmetric quantum mechanics: 
$X^\mu(x)(\mu=5..10)$ and $X^a(x)(a=1..4)$, corresponding to the position of 
the D-particles in the `flat' directions, respectively in the orbifold 
directions. These carry $n\times n$ Chan-Paton matrix indices. General 
considerations \cite{Polchinski} instruct one to keep only the 
$\Z_n$ -invariant part of $ X^\mu,X^a$, where $\Z_n$ acts on $X^\mu$, $X^a$ as:
\bea
X^\mu\rightarrow \rho_{reg}(\alpha)X^\mu\rho_{reg}(\alpha)^{-1} \nn \\
X^a\rightarrow \rho_f(\alpha) \rho_{reg}(\alpha) X^a \rho_{reg}(\alpha)^{-1}~~
\nn .
\eea
The extra-factor $\rho_f(\alpha)$ in the action on $X^a$ reflects the 
nontrivial transformation law of the coordinates parallel to the orbifold 
directions. The compact part of moduli space of vacua coincides 
\cite{DM} with the orbifold $\C^2/\Z_n$ itself. 
More general actions on the Chan-Paton 
factors are possible and were discussed in some 
detail in \cite{DM}. Some of these do not have a standard D-particle 
interpretation, corresponding instead to bound states containing fractional 
numbers of D-particles and/or D2-branes \cite{fractional_D2}. 
In general, it turns out \cite{DM,JM} that the 
moduli space of the resulting quantum mechanics is given by a singular 
hyperkahler quotient of matrix data.

A second application of the quotienting procedure was proposed in 
\cite{Taylor} for the case $G=\Z^4$. This leads to compactifications 
of D-particle systems on a four-torus $T=\R^4/\Z^4$. 
Once again, the theory of $r$ D-particles on $T$ is obtained by considering 
their system of images on $\R^4$ under the $\Z^4$-action. 
This time, however, the number of images is 
infinite, which leads to an effective description given by a supersymmetric 
$U(r)$ gauge field theory defined on the dual torus $T'$. The images are 
distributed according to the action of $\Z^4$ on $\R^4$, but 
there is again considerable freedom in the choice of the action on the 
Chan-Paton factors. This Chan-Paton 
representation is reflected in the topology of the bundle over $T'$ 
in which the effective field theory is defined. In fact, Taylor's original 
approach takes the Chan-Paton representation to be the regular 
representation of $\Z^4$, which leads to a trivial 
bundle over $T'$, but more general representations can be considered, and 
they lead to nontrivial bundles. Such compactifications can be interpreted as 
systems of D-particles over $T$ which also contain D2 and D4-branes. 
In general, the moduli space of supersymmetric vacua is given 
by the moduli space of instantons in the bundle. In the case of the 
trivial bundle, this is just the moduli space of flat connections. 
Again the simplest case is that of one D-particle, 
which leads to a $U(1)$ gauge theory and to the moduli space of flat 
line bundles over $T'$, which coincides --- as expected --- with the original 
torus $T$. 

Note that the situation in the compact case is more involved because
of the presence of an infinity of images. While in the noncompact 
case there exists a quantum-mechanical description of the moduli 
space, in the compact case only a field-theoretic description is 
available. The gauge symmetry is infinite dimensional for the
latter.

Viewed more abstractly, both the construction of \cite{DM} and 
that of \cite{Taylor} are procedures for quotienting the effective theory 
on the covering space by a discrete group--finite, in the first case, and 
infinite in the second. In general, then, one can follow the same pattern 
for describing D-particles over $\R^4/G$ for any discrete group $G$. 
Our focus in this paper will be on the case of the semidirect product 
of $\Z^4$ and $\Z_n$, with appropriate integer $n$, which leads to a 
theory of D-particles on $T^4/\Z_n$. In fact, we will be primarily
concerned with the case of one D-particle only. 
This can be described in a way which is very similar to that of 
\cite{Taylor}, by considering an image of the D-particle for each 
element of $G$, as we will explain in detail in the next section. 
However, it is by now well  understood that another and particularly useful 
way of formulating this sort of problem is via the more abstract approach of 
\cite{CDS}. This starts with a supersymmetric `quantum mechanics' of variables 
$X^\mu, {\cal X}^a$ valued in a Hilbert space ${\cal H}$ and imposes 
projection 
conditions with respect to a unitary representation $U$ of $G$ in ${\cal H}$. 
The freedom in the choice of $U$ corresponds essentially 
to the freedom in the choice of 
the action of $G$ on the Chan-Paton factors. Then the case of 
finite-dimensional ${\cal H}$ allows us to describe quotients $\R^4/G$ 
for a finite group $G$, while the case of an infinite (but discrete) $G$ 
can be described by taking ${\cal H}$ to be a (separable) 
infinite-dimensional Hilbert space\footnote{Moreover, if one replaces 
$U$ by a {\em projective} representation of $G$, then one obtains 
noncommutative-geometric compactifications as in \cite{CDS}. In this paper, 
however, we will only consider proper representations.}. 
This approach has 
major advantages in terms of generality and clarity as well as in terms 
of computational power. As we will show explicitly in the next section, 
its relation 
to the more intutive approach of \cite{Taylor} is similar to the relation 
between the abstract operator formalism of quantum mechanics and its 
formulation in a particular representation: Taylor's description appears 
simply by chosing a particular pair of bases of 
the Hilbert space ${\cal H}$: 
a basis leading to the D-particle representation and a basis leading to the 
D4-brane representation. From this point of view, the Fourier transform of 
\cite{Taylor} is simply the change of basis between these, 
while the compactification procedure of \cite{CDS} is the overarching 
`operator formulation'.  

\subsection{The resolution mechanism in the noncompact case}

In the case of $\C^2/\Gamma$, it was shown in \cite{DM} that the effective 
quantum mechanics describing the D-particle system admits Fayet-Iliopoulos
terms. More precisely, one can turn on such a term for each central 
generator of the effective gauge group. (The gauge group is generated by 
the unitary gauge transformations that commute with the orbifold 
projection, while the effective gauge group is obtained by modding 
out a $U(1)$ factor which acts trivially on $X^\mu,X^a$.) 
The moduli space of the resulting theory 
gives a resolution of the orbifold, and, in the Matrix theory limit at 
least, this can be interpreted as a resolution of the space itself. 
In fact, the quantum-mechanical description gives the resolved moduli space 
as a hyperkahler quotient, in a form which is {\em identical} to the 
hyperkahler quotient description given in \cite{K} to the minimal resolution 
of the orbifold singularity. The resolved space $X$ is a smooth hyperkahler 
manifold which is asymptotically locally euclidean (i.e. it is an ALE space). 

For compact orbifolds, it is not immediately clear how such a procedure 
would be applied. Indeed, what we will need is some analogue of the 
resolution mechanism of \cite{DM}, given that fundmental degrees of 
freedom are not just finite-dimensional matrices but equivariant
and intrinsically nonabelian connections. By analogy with the process 
of \cite{DM}, we could hope that, starting with the infinite lattice
of D-particles on $R^4$, one finds a {\em natural} mechanism for producing a 
deformation of orbifold gauge field theory  --- {\em still defined 
over $T'/\Z_n$} --- whose vacuum moduli space gives a resolution of 
the moduli space of the original system.

\section{Formulation of $T^4/{\bf Z_n}$ D-particle orbifolds }

\subsection{The abstract set-up}

To obtain a compactification preserving half of the supersymmetry 
of the type IIA 
theory, we start with a two-dimensional hermitian vector space $V$ and 
a maximal rank lattice $\Lambda \subset V$. We let  $<,>$ denote the 
hermitian scalar 
product on $V$ and $(,):=Re<,>$ the associated euclidean product on the 
underlying real vector space $V_R \approx \R^4$. If  
$\Lambda':=\{x \in V | (x, t) \in \Z
, ~\forall t \in \Lambda\}$ is the dual 
lattice, we have dual 
4-tori $T=V/\Lambda$ and  $T'=V/\Lambda'$. We consider a faithful 
representation $\gamma:\Z_n \rightarrow SU(V)$ of $\Z_n$ by {\em special} 
unitary transformations of $V$. 
To have a well-defined quotient, we 
must assume that $\Lambda$ is invariant with respect to this action:
\bdm
\gamma(\alpha)(\Lambda)=\Lambda ~~,
\edm
for all $\alpha \in \Z_n$ 
\footnote{Note that $V$ admits a unique quaternionic 
structure compatible both with its original complex structure and its 
hermitian metric. With respect to this quaternionic structure, $SU(V)$ is 
identified with the group of unitary symplectic automorphisms of $V$ and 
$\gamma$ acts by elements of $USp(V)$.   
Therefore, $T$ and $T'$ can be viewed as flat quaternionic manifolds 
in a natural way. The reason for requiring $\gamma$ to act by special 
unitary transformations is that we wish to preserve this quaternionic 
structure, which assures us that orbifolding by $\Z_n$ breaks exactly half 
of supersymmetry. Alternatively, we wish to have a Calabi-Yau orbifold, 
so that $\gamma$ must preserve triviality of the canonical line bundle, and 
this requires ${\rm det}\gamma(\alpha)=1$. Examples of lattices invariant 
under such $\Z_n$ actions are easily constructed. The simplest case is  
the action of $\Z_2$ by negation on $V$, which preserves any lattice 
$\Lambda$. Another is the case of the `fundamental' representation of $\Z_4$, 
defined by $\gamma_f(\alpha)={\rm diag}(i^\alpha,i^{-\alpha})$; lattices 
invariant under this action can be obtained by taking 
$\Lambda:=(\Z+i\Z)\epsilon_1+
(\Z+i\Z)\epsilon_2$, with $(\epsilon_1,\epsilon_2)$ the complex basis of $V$
which diagonalizes $\gamma_f$.  Although the quaternionic structure 
will be important in section 4, we will present our 
discussion in the simpler language of hermitian vector spaces. 
This amounts to singling out a preferred complex structure.}. 

Consider the group  $G$ of real affine transformations of $V$ generated by 
translations with $t \in \Lambda$ and by the linear transformations  
$\gamma(\alpha) (\alpha \in \Z_n)$. This is given by actions 
$v \rightarrow \gamma(\alpha)v +t$ on the vectors $v$ of $V$. 
As an abstract group, $G$ is a semidirect product of $\Lambda$ and $\Z_n$, 
having $\Lambda$ as a normal subgroup. Its composition law is given by:
\bdm
(t',\alpha')\cdot (t,\alpha)=(\gamma(\alpha')t+t',\alpha+\alpha')
\edm
and the inversion by $(t,\alpha)^{-1}=(-\gamma(-\alpha)t,-\alpha)$. 
Here $\alpha \in \Z_n$ and $t \in \Lambda$. 

To formulate the orbifold problem along the lines of \cite{CDS}, 
we consider a Hilbert space 
${\cal H}$, a unitary representation $U$ of $G$ in ${\cal H}$ and 
selfadjoint operators $X^\mu, {\cal X}^a$ in ${\cal H}$, subject to the 
`projection conditions':
\be 
\label{proj1}
U(t, \alpha)^{-1}X^\mu U(t, \alpha)=X^\mu
\ee
\be
\label{proj2}
U(t,\alpha)^{-1}X U(t,\alpha)=\gamma(\alpha)X + 2 \pi t ~Id_{\cal H}~~.
\ee
where  $X= {\cal X}^a \otimes e_a$ is a $V$-valued operator, with $e_1..e_4$ a 
real basis of $V$. In ~(\ref{proj2}), $U(t,\alpha)$ acts ${\cal X}^a$, while 
$\gamma(\alpha)$ acts on the basis elements $e_a$.  
Physically, $X^\mu, {\cal X}^a$ 
describe `infinite by infinite' Chan-Paton matrices, while $U$ gives the 
action of the group $G$ on the Chan-Paton factors. Therefore, the abstract 
projection conditions above are nothing else than the adaptation of the usual 
orbifold projection conditions on open strings \cite{Polchinski} to the case 
of the infinite group $G$. 

\subsection{The D-particle basis}

To describe one D-particle on $T/\Z_n$, we choose 
$U$ to be the regular representation of $G$. This is characterized 
up to unitary equivalence by the requirement that 
there exist a vector  $h \in {\cal H}$ such that the set $
\{|t,\alpha>:=U(t,\alpha)h | t \in \Lambda, \alpha \in \Z_n \}$ is 
an orthonormal basis of ${\cal H}$. In this case:
\bdm
U(t',\alpha')|t,\alpha>=|\gamma(\alpha')t+t',\alpha +\alpha'>=
|(t',\alpha')\cdot (t,\alpha)>~~,
\edm
so that, defining 
$X^\mu_{s,\alpha;t,\beta}:=<s,\alpha|X^\mu|t,\beta> \in \C$ and 
$X_{s,\alpha;t,\beta}:=<s,\alpha|X|t,\beta> \in V$, the projection conditions 
become:
\bea
X^\mu_{(t',\alpha')\cdot(s,\alpha);(t',\alpha')\cdot(t,\beta)}=
X^\mu_{s,\alpha;t,\beta} \nn\\
X_{(t',\alpha')\cdot(s,\alpha);(t',\alpha')\cdot(t,\beta)}=
\gamma(\alpha')\cdot X_{s,\alpha;t,\beta}+
2\pi t' \delta_{\alpha,\beta}\delta_{s,t} \nn~~.
\eea
As the states of the above basis are labelled by the group elements 
$(t,\alpha)$, we can think of our Hilbert space as the space of 
square-summable sequences $\phi:\Lambda\bowtie \Z_n \rightarrow \C$, 
$\sum_{t,\alpha}{|\phi(t,\alpha)|^2}< \infty $. Then  
the group $G$ acts by 
$[U(t',\alpha')\phi](t,\alpha)
=\phi((t',\alpha')^{-1}\cdot (t,\alpha))$, which is 
the defining property of the regular representation. The abstract 
variables $X^\mu,X$ become infinite 
matrices $X^\mu_{s,\alpha;t,\beta},X_{s,\alpha;t,\beta}$ with entries in $\C$, 
respectively $V$. Their physical interpretation is as Chan-Paton matrices 
of open strings connecting pairs of D-particles $(s,\alpha)$ and 
$(t,\beta)$. We can interpret the D-particle 
indexed by $(t,\alpha)$ as the image by the element $(t,\alpha) \in G$ of the 
D-particle $(0,0)$. This is the direct generalization of the representation 
used in \cite{Taylor} for the case of one D-particle on a torus, and we 
call 
it the `D-particle' representation. A two-dimensional model of the situation 
is drawn below (for the case $n=2$). 

\

\iffigs
\begin{figure}[htpb]
\begin{center}
\input{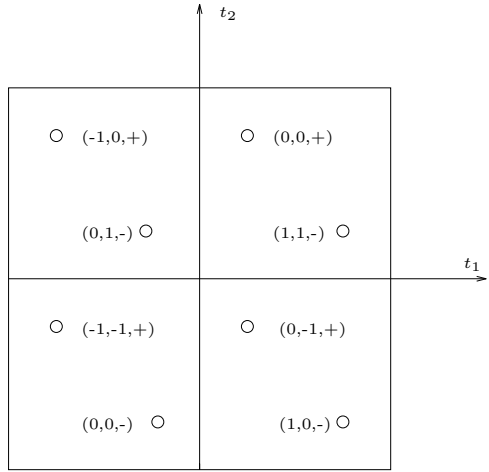}
\caption{\footnotesize A two-dimensional model of one 
D-particle on the covering space together with 
some of its images. For clarity we denoted the elements ${\hat 0}, 
{\hat 1}$ of $\Z_2$ by $+={\hat 0}$ and $-={\hat 1}$. The first two 
coordinates give the lattice translation vector $t=(t_1,t_2)$. All images 
are obtained by acting with various $(t_1,t_2,\alpha)$ on $(0,0,+)$.}
\label{Figure1}
\end{center}
\end{figure}
\fi

\

\subsection{The `intermediate' basis}

The action of the $\Z_n$ subgroup of $G$ can be diagonalized by performing 
a discrete Fourier transform. For this, let $\xi:=e^{\frac{2\pi i}{n}}$. 
Since:
\bdm
\sum_{\alpha \in \Z_n}{\xi^{\alpha \beta}}=n \delta_{\beta,0}
\edm
(where $\delta_{\beta,0}$ is taken in $\Z_n$, i.e. 
$\delta_{n,0}=\delta_{0,0}=1$), we can define a new 
orthonormal basis of ${\cal H}$ by:
\be
\label{intermediate}
|t,\alpha):=\frac{1}{\sqrt{n}}\sum_{\beta \in \Z_n}
{\xi^{\alpha\beta}|t,\beta>}~~.
\ee
In this basis, the action of $G$ becomes:
\bdm
U(t',\alpha')|t,\alpha)=\xi^{-\alpha\alpha'}
|\gamma(\alpha')t+t',\alpha)~~.
\edm
We call this the `intermediate' representation. The discrete Fourier 
transform from the D-particle representation to the intermediate 
representation 
is similar to the transform effected in \cite{DM,DGM} in order to diagonalize 
the $\Z_n$ action on the Chan-Paton factors. This representation has an 
interpretation similar to that of the D-particle basis 
and is further discussed in the appendix. 

\subsection{The D4-brane basis}

The $T$-dual description of our system can be obtained by a further change of 
basis of ${\cal H}$\footnote{We consider the 
Fourier transform from the space of sequences enumerated by $\Lambda$ to the 
space of distributions on the dual torus $T'=V/\Lambda'$. 
For each $t \in \Lambda$, we have a function $e^{2\pi i (x,t)}$ on 
$V$. This is $\Lambda'$-periodic 
and induces a function 
on $T'$, which we still denote  
by $e^{2\pi i (x,t)}$. To define the Fourier transform we 
consider the natural measure 
$d\mu_{T'}(x):=[dx]$ on $T'$, induced from the Lesbegue measure on the 
covering space and normalized such that the elementary cells of $\Lambda'$ 
have volume $1$. 
The Fourier coefficients  of a function $f$ on $T'$ are then 
given by:
\bdm
f_t:=\int_{T'}{[dx]f(x)e^{2 \pi i(x,t)}} ~(t \in \Lambda)~~, 
\edm
with the inversion formula:
\bdm
f(x)=\sum_{t \in \Lambda}f_t e^{-2 \pi i(x,t)}~~.
\edm
This extends to distributions on $T'$, giving the 
completeness and orthogonality relations :
\bea
\sum_{t \in \Lambda}e^{2\pi i (x-x',t)}=\delta_{T'}(x-x') ~~;~~&&
\int_{T'}{[dx]e^{2 \pi i (x,t-t')}=\delta_{t,t'}}\nn
\eea
Here the difference $x-x'$ is understood as an operation in the group 
$(T',+)$ (the abelian group of points of $T'$ with the natural addition 
induced from $V$ by the projection $V\rightarrow T'=V/\Lambda'$), while  
$\delta_{T'}(x)$ is the Dirac distribution on $T'$, 
normalized with respect to the above measure:
\bdm
\int_{T'}{[dx]\delta_{T'}(x)}=1~~.
\edm}
:
\be
\label{fourier_abstract}
|x,\alpha>:=\sum_{t\in \Lambda}
{e^{2\pi i (x,t)}|t, \alpha)},
\ee
for $x \in T'$ and $\alpha \in \Z_n$.  
We have the orthogonality and completeness conditions:
\bea 
<x,\alpha|y,\beta>=\delta_{T'}(x-y)\delta_{\alpha,\beta} \nn \\
\int_{T'}{[dx]\sum_{\alpha \in \Z_n}{|x,\alpha><x,\alpha|}}=Id_{\cal H}\nn ~~.
\eea

In this `D4-brane' representation, ${\cal H}$ is the Hilbert space 
$L^2(T')\otimes \C^n$, and the action of $G$ becomes:  
\bdm
U(t',\alpha')|x,\alpha>=\xi^{-\alpha \alpha'}
e^{-2 \pi i (\gamma(\alpha')x,t')} |\gamma(\alpha')x, \alpha>~~
\edm
(here we used unitarity of $\gamma(\alpha')$ and the fact that $(,)=Re<,>$).
This shows that $U(t,0)$ has (generalized) eigenvalues 
$e^{-2 \pi i (x,t)}$ 
with generalized eigenspaces 
$I_x:={\rm Span}((|x\alpha>)_{\alpha \in \Z_n})$. 
Since condition ~(\ref{proj1}) requires $X^\mu$ to preserve these 
subspaces, we must have:
\bdm
X^\mu|x,\alpha>=X^\mu_{\beta,\alpha}(x)|x,\beta> ~.
\edm
Hence $X^\mu$ become multiplication operators by the hermitian 
matrices $X^\mu(x):=(X^\mu_{\alpha,\beta}(x))_{\alpha,\beta \in \Z_n}$. 
Then taking $t=0$ in ~(\ref{proj1}) gives:
\bdm
X^\mu_{\alpha,\beta}(\gamma(\alpha')x)=
\xi^{-(\alpha-\beta)\alpha'}X^\mu_{\alpha,\beta}(x)~~.
\edm

\subsection{The D4-brane orbifold gauge theory}

\

\noindent{\bf Orbifold connections}

\

To solve ~(\ref{proj2}), define the operator 
$P$ by $P|t,\alpha)= 2\pi t |t,\alpha)$ and let 
$X:=P+A$. Since $P$ satisfies 
$U(t',\alpha')^{-1}PU(t',\alpha')=\gamma(\alpha')P+2\pi t'$, 
we can rewrite ~(\ref{proj2}) as:
\be
\label{proj2'}
U(t',\alpha')^{-1}AU(t',\alpha')=\gamma(\alpha')A
\ee 
Taking $\alpha'=0$ shows that $A$ must have the form:
\bdm
A|x,\alpha>=A_{\beta,\alpha}(x)|x,\beta> ~,
\edm
which gives a matrix multiplication operator 
$A(x)=(A_{\alpha,\beta}(x))_{\alpha,\beta \in \Z_n}$ (with entries in $V$). 
Then ~(\ref{proj2'}) becomes:
\be
\label{proj2''}
A_{\alpha,\beta}(\gamma(\alpha')x)= 
\xi^{-(\alpha-\beta)\alpha'}\gamma(\alpha')A_{\alpha,\beta}(x)
\ee
Since in the D4-brane representation we have $P=+i\nabla_x$, with 
$\nabla$ the gradient associated to the metric induced by $(,)$ on $T'$,
we obtain:
\bdm
X=i[\nabla_x -iA(x)]~~.
\edm
To formulate this more geometrically, pick any real basis $e_a(a=1..4)$ 
of $V$ and expand:
\bea
A(x)=\sum_{a=1..4}{{\cal A}^a(x)e_a}~~\nn\\
X=\sum_{a=1..4}{{\cal X}^a e_a} \nn~~.
\eea
Defining:
\bdm
{\cal A}_a(x):=g_{ab}{\cal A}^b(x)~~
\edm
(with $g_{ab}=(e_a,e_b)$) and lowering indices in (\ref{proj2''}) we obtain:
\bdm
{\cal A}_a(\gamma(\alpha')x)=\rho_{reg}(-\alpha')\gamma(-\alpha')^b_{~a}
{\cal A}_b(x)\rho_{reg}(\alpha')~~,
\edm
where we again used unitarity of $\gamma(\alpha)$, the fact that 
$(,)=Re<,>$ and we let  
$\rho_{reg}$ denote the regular matrix representation of $\Z_n$: 
$\rho_{reg}(\alpha)=
{\rm diag}(1,\xi^\alpha,\xi^{2\alpha}...\xi^{(n-1)\alpha})$. 
Replacing $\alpha'$ by $-\alpha$, we can rewrite this as:
\bdm
(\gamma(-\alpha)^t)_a^{~b}{\cal A}_b(\gamma(-\alpha)x)=\rho_{reg}(\alpha)
{\cal A}_a(x)\rho_{reg}(-\alpha)~~.
\edm
Defining the matrix-valued one-form $A=\sum_{a=1..4}{{\cal A}_a(x)dx^a}$
\footnote{From now on, $A$ will denote this one-form and not the object 
$A(x) \in V$ considered previously. We hope that this does not produce 
any confusion.}, 
the last relation becomes:
\be
\label{projection}
\gamma(-\alpha)^*A=\rho_{reg}(\alpha)A\rho_{reg}(\alpha)^{-1}~~,
\ee
where $\gamma(-\alpha)^*$ denotes the pull-back. 
$A$ can be viewed as a connection 
in a rank $n$ hermitian bundle $I$ over $T'$, 
whose fiber over $x$ is given by the vector space 
$I_x={\rm Span}(|x\alpha>)_{\alpha \in \Z_n}$. 
The elements $|x\alpha>$ are then  
identified with sections $s_\alpha(x)$ of $I$. This gives a global 
orthonormal frame $(s_\alpha)_{\alpha \in \Z_n}$ of $I$, which shows that 
$I$ is the trivial rank $n$ bundle. Then $A$ is a connection matrix 
in this frame, with associated covariant derivative operator:
\bdm
-i{\cal X}_a=-ig_{ab}{\cal X}^b=\partial_a-i{\cal A}_a(x)~~.
\edm 

The conclusion is that $-i{\cal X}_a$ becomes 
the covariant derivative of a $U(n)$ gauge connection 
on a trivial bundle over $T'$, 
subject to the projection condition (\ref{projection}). 
This requires the connection 
one-form to be invariant under the action of the 
orbifold group, up to the constant gauge transformation $\rho_{reg}(\alpha)$.  
This type of condition is  natural when defining connections  
on an orbifold: the projection condition specifies the way in which the 
connection `twists' around the fixed point. The gauge 
transformations associated to $\alpha \in \Z_n$ 
induce identifications between the fibers of the underlying 
bundle, which make it into a $V$-bundle. The identifications describing the 
$V$-bundle structure can be used to induce a natural action of the orbifold 
group on the space of sections.
Then (\ref{projection}) is 
equivalent to the condition that the covariant differentiation operator be 
equivariant with respect to this action on sections, which can be interpreted 
as a compatibility condition between the connection and the $V$-structure.  
Thus, the mathematical meaning of (\ref{projection}) is that $A$ is an 
equivariant connection in the $V$-bundle associated to $I$ and to the 
identifications induced by $\rho_{reg}(\alpha)$.
Since $I$ is the trivial rank $n$ bundle 
over $T'$ and the gauge 
transformations $\rho_{reg}(\alpha)$ defining the $V$-structure 
are constant over $T'$, we have a `product' $V$-bundle structure 
$E=T'\times_{\Z_n}(\C^n,\rho_{reg})$ 
associated to the regular representation of $\Z_n$. 

\

\noindent{\bf Orbifold gauge transformations}

\

Let us discuss the equivalence relation needed to build the 
moduli space. Clearly we should identify two representations 
$(X^\mu,X,U)$ and $(X'^\mu,X',U')$ in ${\cal H}$ 
if there exists a {\em unitary} operator $U$ in ${\cal H}$   
such that $U$ commutes with all 
$U(t,\alpha)$ and intertwines $X^\mu,X$ and $X^{'\mu},X'$.  
The group of such operators $U$ is the  `projected symmetry group'. 
In the D4-brane basis, the condition that $U$ commutes with all 
$U(t,\alpha)$ constrains 
$U$ to be a local $U(n)$-valued multiplication operator $U(x)$, 
satisfying the projection:
\be
\label{gauge_projection}
U(\gamma(-\alpha)(x))= \rho_{reg}(\alpha) U(x) \rho_{reg}(\alpha)^{-1}~~.
\ee
The action $X\rightarrow UXU^{-1}$ 
of $U$ on $X$ is reflected by the usual gauge transformation law 
\bdm
A\rightarrow UAU^{-1}-idUU^{-1}
\edm
while the action $X^\mu\rightarrow UX^\mu U^{-1}$ on $X^\mu$ becomes:
\bdm
X^\mu(x)\rightarrow U(x)X^\mu(x)U(x)^{-1}~~.
\edm
Thus, we can view $X^\mu(x)$ as scalar fields on $T'$ in the adjoint 
representation of $U(n)$.

Mathematically, (\ref{gauge_projection}) means that  
$U$ defines a `$V$-gauge transformation' of $E$, i.e. a unitary bundle 
automorphism which commutes with the $V$-structure.  
Thus, we should identify two $V$-connections $A,A'$ if they are 
$V$-gauge equivalent in the $V$-bundle $E$.  

\

\noindent{\bf The compact part of the moduli space}

\

To construct the physically interesting moduli space we must also impose 
the vacuum conditions stemming from the standard super-Yang-Mills action. 
Considering a real basis $e_1..e_4$ of $V$ as before 
and defining $g_{ab}:=(e_a,e_b)$
($a,b=1..4$), we can write the bosonic part of the action as
\footnote{We will allways use $Tr$ to denote the `total' trace in the Hilbert 
space ${\cal H}$ and $tr$ to denote the matrix trace in $u(2)$.}:
\begin{eqnarray*}
\label{action}
\lefteqn{S=\frac{1}{2\rho g}\int{dt~\{Tr({\dot X^\mu}{\dot X^\mu})+
g_{ab}Tr~({\dot {\cal X}^a}{\dot {\cal X}^b})\}. }} \\
&& +\frac{1}{4\rho g}\int{dt~\{Tr~([X^\mu,X^\nu][X^\mu,X^\nu])+
2g_{ab}Tr~([X^\mu,{\cal X}^a][X^\mu,{\cal X}^b])\}} \\  
&&+ \frac{1}{4\rho g}~\int{dt~\{g_{ac}g_{bd}Tr~([{\cal X}^a,{\cal X}^b],
[{\cal X}^c,{\cal X}^d])\}}~~.
\end{eqnarray*}
Here $\rho$ is a normalization factor, which is needed to eliminate the 
`infinite measure' of the $\Z^4$ symmetry group of the projected theory. 
Formally, $\rho$ is given by $\rho=\sum_{t \in \Lambda}{1}=\delta_{T'}(0)$. 
In this paper, we will treat the issue of such infinities quite formally, 
avoiding a serious discussion of regularization procedures.

The vacuum constraints are :
\be
\label{w1}
[X^\mu,X^\nu]=0 
\ee
\be
\label{w2}
[X^\mu,{\cal X}^a]=0
\ee
\be
\label{w3}
[{\cal X}^a,{\cal X}^b]=0~~.
\ee
In the D4-brane basis, these become:
\be
\label{c1}
[X^\mu(x),X^\nu(x)]=0
\ee
\be
\label{c2}
{\cal D}^{(adj)}_aX^\mu =\partial_a X^\mu(x) -i[{\cal A}_a(x),X^\mu(x)]=0
\ee
\be
\label{c3}
{\cal F}_{ab}= \partial_a {\cal A}_b-\partial_b {\cal A}_a-
i[{\cal A}_a,{\cal A}_b]=0~~,
\ee
where ${\cal D}^{(adj)}$ is the covariant derivative associated to $A$ in 
the adjoint representation. 

The adjoint scalars $X^\mu(x)$ are sections of a trivial 
Lie agebra bundle (of type $u(n)$) 
over $T'$; their modes describe the oscillations of the D4-brane in 
the uncompactified directions. The full system of constraints is a coupled 
nonlinear system, somewhat similar to the Yang-Mills-Higgs equations. 
One can obtain a simplification by assuming 
that $X^\mu(x)$ are slowly varying over $T'$. Under this adiabatic 
approximation the total moduli space becomes a fibration over the space of 
constant matrices $X^\mu\in u(n)$ satisfying $~(\ref{w1})$, 
whose fibre is the moduli 
space of solutions to $~(\ref{c2})$ and $~(\ref{c3})$. Then the D-particle 
has a well-defined position in the uncompactified space
directions (specified by the 
common eigenvalues of $X^\mu$), so 
that the brany-geometry of the uncompactified part of the space reduces to its 
classical geometry. This is the limit we are interested in in this paper, 
since we want to probe the brany-geometry of the compact part of the space 
only. To further simplify the problem we will take 
$X^\mu=0,~\forall \mu $. 
Then {\em the desired 
moduli space is given by the moduli space  
of flat $V$-connections in the $V$-bundle $E$, 
modulo $V$-gauge transformations}.

It is expected that this moduli space coincides with the original orbifold 
$T/\Z_n$. With the above formulation, we wish to present a clear proof
of this claim. We will proceed by considering an equivariant form of 
the argument which relates flat connections to their holonomy. Further,
to simplify the presentation, we will only consider the case $n=2$, which 
is the focus of the rest of this paper, but similar arguments can be applied 
for other cases. We will also restrict to $SU(2)$ connections, 
which amounts to imposing the tracelessness condition on ${\cal A}_a$ at 
every point of $T'$. (General $U(n)$ connections can be studied by similar 
methods.)

\section{The singular $SU(2)$ moduli space for the case $n=2$}

When $n=2$ we let $\Z_2$ act on $V$ by negation $x \rightarrow -x, 
~\forall x \in V$. Any lattice in $V$ is invariant 
under this action, which descends to the usual negation of the complex torus 
$T'=V/\Lambda'$. It is well-known \cite{LB,Mumford} that the negation of 
$T'$ has 16 fixed points. These 
form a subgroup of $(T',+)$ which is isomorphic
\footnote{The fixed points $x_l$ are the projections modulo $\Lambda'$ 
of some points $m_l$ of the covering space $V$. The fixed-point condition 
$x_l=-x_l \Leftrightarrow 2x_l=0 ~(\rm {in ~}(T',+))$ is equivalent to 
the constraint $m_l \in \frac{1}{2}\Lambda'$. Thus the fixed points are 
in bijection with the elements of the quotient group 
$(\frac{1}{2}\Lambda')/\Lambda'\approx (\Z_2)^4$.} with $(\Z_2)^4$.

The regular representation of $\Z_2$ is generated by the Pauli matrix 
$\rho_{reg}({\hat 1})=\sigma_3$. The projection condition on the 
connection becomes:
\be
\label{A_proj}
\sigma_3{\cal A}_a(x)\sigma_3=-{\cal A}_a(-x)~~,
\ee
while the projection condition on gauge transformations is:
\bdm
\sigma_3U(x)\sigma_3=U(-x)~~.
\edm
These restrict ${\cal A}_a$, $U$ to the forms:
\bea
{\cal A}_a(x)=\left(\begin{array}{cc}
	o_1(x) & e_1(x)\\
	e_2(x) & o_2(x) 
	\end{array}\right)~~, \nn\\
U(x)=\left(\begin{array}{cc}
	e_1(x) & o_1(x)\\
	o_2(x) & e_2(x) 
	\end{array}\right)~~, \nn\\
\eea
where $e_k,o_k$ denote even, respectively odd complex-valued 
functions defined over $T'$. It is important to realize that, in the 
above relations, $-x$ represents the negation on $T'$ (the opposite of the 
element $x$ in the abelian group $(T',+)$). Thus, a point $x \in T'$
is a solution of the equation $x=-x \Leftrightarrow 2x=0$ 
if and only if it is a fixed point of the 
$\Z_2$ action. We will denote the fixed points by $x_0...x_{15}$. 

\subsection{Characterization of the moduli space}

Choose a fixed global trivialization of the underlying bundle $I$ of $E$  
over $T'$. Then any $su(2)$ 
connection $A$ on $I$ is represented by a globally defined
$su(2)$-valued 1-form over $T'$. Any such $1$-form is induced by 
a periodic $1$-form $A(v)$ defined on the covering space $V$. Here 
periodicity means :
\bdm
\tau_t^*(A)=A, ~\forall t \in \Lambda'
\edm
where $\tau_t(v):=v+t$ ($v \in V, t \in \Lambda'$) is the translation by $t$ 
on $V$. If one choses an arbitrary real basis $e_1. .e_4$ of $V$, 
in which $A$ has components ${\cal A}_a$, then periodicity reads:
\bdm
{\cal A}_a(v+t)={\cal A}_a(v), ~\forall t \in \Lambda'~~.
\edm

\

\noindent{\bf Reduction to parallel transport operators}

\

Clearly $A(x)$ is flat on $T'$ if and only if 
$A(v)$ is flat on $V$. In this case, 
the parallel transport of $A(v)$ gives a well-defined $SU(2)$ matrix-valued 
function $P(v)$ on $V$
(depending on our trivialization of $I$). We choose $0\in V$
(with projection $x_0=O \in T'$) as a base point for defining $P(v)$.
Then $P(v)$ satisfies:
\bea
dP(v)=iA(v)P(v)\nn\\
P(0)=1~~. \nn
\eea
Together with $\Lambda'$-periodicity of $A(v)$, this equation immediately 
leads to the property:
\be
\label{quasiper}
P(v+t)=P(v)P(t), ~\forall v \in V, ~\forall t \in \Lambda'
\ee
(that is, $P(v)$ is $\Lambda'$-`quasiperiodic'). 
It is easy to see that the $\Z_2$ projection condition on $A$ implies: 
\be
\label{pt_proj}
P(-v)=\sigma_3 P(v) \sigma_3~~.
\ee

The gauge transformations $U(x)$ of $A(x)$ are induced by $\Lambda'$-periodic 
$SU(2)$-valued functions $U(v)$ on $V$. They act on $P(v)$ by :
\be
\label{pt_gauge}
P(v) \rightarrow U(v)P(v)U(0)^{-1}~~.
\ee
Moreover, the projection condition on $U(x)$ is equivalent to :
\be
\label{gauge_proj}
U(-v)=\sigma_3 U(v) \sigma_3~~.
\ee

Conversely, if one is given a map $P:V \rightarrow SU(2)$ satisfying 
~(\ref{quasiper}) and (\ref{pt_proj}), then $A(v):=-i(dP(v))P(v)^{-1}$ is a 
$\Lambda'$-periodic 1-form 
on $V$ which induces a flat $SU(2)$ connection $A(x)$ on $T'$ satisfying the 
$\Z_2$ projection condition (\ref{A_proj}). Moreover, if $U(v)$ is a periodic 
$su(2)$-valued function on $V$ which satisfies (\ref{gauge_proj}), then 
the transformation (\ref{pt_gauge}) of $P$ is equivalent to the gauge 
transformation of $A(x)$ by the associated projected gauge group element 
$U(x)$. Hence the desired moduli space is given by:
\be
\label{moduli_1}
{\cal M}_{flat}=\frac{\{P:V\rightarrow SU(2)|P {\rm ~satisfies~} 
(\ref{quasiper}) {\rm ~and~}(\ref{pt_proj})\}}{
\{P(v)\rightarrow U(v)P(v)U(0)^{-1}| U {\rm ~is~} \Lambda'
{\rm -periodic~and~satisfies~}
(\ref{gauge_proj})\}}~~.
\ee

\

\noindent{\bf Reduction to the holonomy representation}

\

To express the moduli space in terms of finite-dimenional matrices, 
we consider the restriction $S:=P|_{\Lambda'}$ of $P$ to the lattice 
$\Lambda'$. The quasiperiodicity property (\ref{quasiper}) of $P$ implies 
that $S:\Lambda' \rightarrow SU(2)$ is a unitary 
representation of the discrete abelian group $(\Lambda',+)$ (the 
{\em holonomy representation}):
\be
\label{repr}
S(t+t')=S(t)S(t'),~\forall t, t' \in \Lambda'~~.
\ee
The projection condition (\ref{pt_proj}) induces a constraint on 
$S(t)$:
\be
\label{hol_proj}
S(t)^{-1}=\sigma_3S(t)\sigma_3~~.
\ee
On the other hand, the gauge transformations (\ref{pt_gauge}) 
(with a $\Lambda'$-periodic $U(v)$) induce actions on the holonomy:
\be
\label{hol_gauge}
S(t)\rightarrow WS(t)W^{-1}~~,
\ee
where $W:=U(0)\in SU(2)$. The projection condition (\ref{gauge_proj}) implies:
\be
\label{W_proj}
W=\sigma_3W\sigma_3~~.
\ee

Thus, we have a well-defined map from ${\cal M}_{flat}$ as given in 
(\ref{moduli_1}) to the quotient of 
all representations $S:\Lambda' \rightarrow SU(2)$, satisfying 
~(\ref{hol_proj}), divided by conjugations ~(\ref{hol_gauge}), with $W$ 
satisfying ~(\ref{W_proj}). 

If two maps $P(v),P'(v)$, satisfying ~(\ref{quasiper}) and ~(\ref{pt_proj}) 
are such that $S'(t)=WS(t)W^{-1}$, with $W$ satisfying ~(\ref{W_proj}), 
then it is easy to see that the  map $U(v):=P'(v)WP(v)^{-1}$ is 
$\Lambda'$-periodic and satisfies ~(\ref{gauge_proj}). Then $U(0)=W$ and 
we have $P'(v)=U(v)P(v)U(0)^{-1}$, so that $P$ and $P'$ are gauge-equivalent. 
This shows that the above map is one to one.

On the other hand, given any representation $S:\Lambda'\rightarrow SU(2)$ 
which satisfies (\ref{hol_proj}), one can immediately 
 construct a parallel transport operator $P(v)$, 
satisfying (\ref{quasiper},\ref{pt_proj}) and such that $P|_{\Lambda'}=S$
\footnote{This can be done as follows. 
Pick an integral basis $\tau_a (a=1..4)$ of $\Lambda'$ and let 
$S_a:=S(\tau_a)$. Then $S_a^{-1}=\sigma_3S_a\sigma_3$ and $[S_a,S_b]=0$. 
Write $S_a=e^{is_a}$, with $s_a$ mutually commuting hermitian 
matrices satisfying $\sigma_3s_a\sigma_3=-s_a$. Then one can take 
$P(v)=e^{iv^as_a}$, for all $v=v^a\tau_a\in V$. In particular, this gives 
a constant representative ${\cal A}_a=s_a$ of the $V$-gauge equivalence 
class of connections having holonomies $(S_1,...,S_4)$ around the cycles
of $T'$ associated to $\tau_1..\tau_4$, as announced in the note on page 16.}.
It follows that the above map is onto. 

In conclusion, our moduli space is, equivalently, given by:
\be
\label{moduli_2}
{\cal M}_{flat}=\frac{\{S:\Lambda'\rightarrow SU(2)|S {\rm ~satisfies~} 
(\ref{repr}) {\rm ~and~}(\ref{hol_proj})\}}{
\{S(t)\rightarrow WS(t)W^{-1}| W \in SU(2) {\rm ~satisfies~}
(\ref{W_proj})\}}~~.
\ee

\

\noindent{\bf Computation of the moduli space}

\

To make this completely explicit, 
let $\tau_1..\tau_4$ be an integral basis of the lattice 
$\Lambda'$ and $S_a:=S(\tau_a)$. Note that $\tau_a$ correspond 
to a basis of cycles on the 
torus, while $S_a$ give the holonomies of $A$ around those cycles. 
The projection condition 
(\ref{hol_proj}) requires $S_a^{-1}=\sigma_3S_a\sigma_3$, while the 
condition (\ref{repr}) that $S_a$ generate a  
representation of the abelian group $(\Lambda',+)$ is equivalent to the 
requirement that they commute:
\be
\label{commutation}
[S_a,S_b]=0 ~\forall a,b=1..4~~.
\ee
Therefore, the desired moduli space is given very explicitly as follows:
\bdm
{\cal M}_{flat}= 
\frac{\{(S_1,S_2,S_3,S_4)|S_a \in SU(2) ~,~ [S_a,S_b]=0, 
~\forall a,b=1..4 {\rm ~and~} \sigma_3 S_a \sigma_3=S_a^{-1} \}}
{\{(S_a)_{a=1..4}\rightarrow (WS_aW^{-1})_{a=1..4}|
 W\in SU(2) {\rm ~and~} \sigma_3W\sigma_3=W\}}~~.
\edm

A simple matrix computation now shows that ${\cal M}_{flat}$ 
coincides (topologically) with the original orbifold  
$K=T/\Z_2$. 
Indeed, 
the general $SU(2)$ solution of the projection constraint ~(\ref{hol_proj})
is:
\bdm
S_a=\left(\begin{array}{cc}
	u_a & v_a \\
	-v_a^* & u_a
	\end{array}\right)
\edm
with $u_a \in \R, v_a\in \C$ satisfying $u_a^2+|v_a|^2=1$. Writing 
$v_a=e^{i\phi_a}s_a$ (with $s_a \in \R$ positive or negative 
and $\phi_a\in [0,\pi)$), 
the commutation relations ~(\ref{commutation}) are equivalent with:
\bdm
e^{2i(\phi_a-\phi_b)}=1~~,
\edm
which shows that $\phi_a$ are all equal. 
Denoting their common value by $\phi \in [0,\pi)$, 
the general solution for $S_a$ is:
\bdm
S_a:=\left(\begin{array}{cc}
	u_a & e^{i\phi}s_a \\
	-e^{-i\phi}s_a & u_a
	\end{array}\right)~~,
\edm
where $u_a,s_a \in \R$ satisfy $u_a^2+s_a^2=1$ (they parametrize a circle 
$S^1$).

The general $SU(2)$ solution of ~(\ref{W_proj}) is:
\bdm
W(\eta)=\left(\begin{array}{cc}
	e^{i\eta} & 0 \\
	0 & e^{-i\eta}
	\end{array}\right)~~,
\edm
with $\eta \in (-\pi,\pi]$. 

Then $W$ acts on $(S_1..S_4)$ via ~(\ref{hol_gauge}) 
by shifting $\phi$ and possibly inverting the sign of $s_a$:
\bea
\phi\rightarrow \phi' \nn\\
s_a\rightarrow (-1)^r s_a~~, \nn
\eea
where $r\in \Z$ and $\phi' \in [0,\pi)$ are 
defined by $\phi+2\eta=\phi'+r\pi$ 
($u_a$ are unchanged under this transformation). 
If some  $s_a$ is nonzero, then we can perform 
the transform by $W(-\frac{1}{2}\phi)$ to go to a `gauge' in which $\phi=0$. 
Then $W(\frac{1}{2}\pi)=W(-\frac{1}{2}\pi)=-1_2$ gives a residual 
identification $s_a\rightarrow -s_a$. If all 
$s_a$ are zero, then $u_a=\pm 1$ and 
we have a fixed point under the action of $W$. 
Thus, we obtain a quotient of the form $(S^1)^4/\Z_2$, where the $a$-th $S^1$ 
is parametrized by $(u_a,s_a)$ and $\Z_2$ acts by $s_a \rightarrow -s_a$. 
This is topologically the same as $T/\Z_2$, as one can see from the 
following two-dimensional model.

\

\iffigs
\begin{figure}[htpb]
\begin{center}
\input{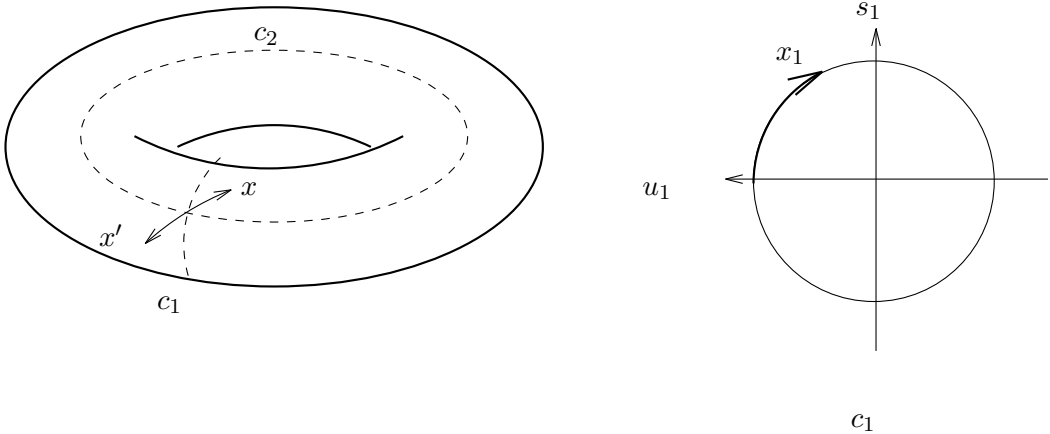}
\caption{\footnotesize Two circles $c_1,c_2$ on a torus associated to a basis 
$\tau_1,\tau_2$
of the lattice $\Lambda$ (in the two-dimensional case).
A point $x=(x_1\tau_1+x_2\tau_2)({\rm mod~} \Lambda')$ 
is shown together with its $\Z_2$ image 
$x'=((-x_1)\tau_1+(-x_2)\tau_2)({\rm mod~}\Lambda')$.  
The right part of the figure shows our coordinates $s_1$, $u_1$
for the first circle. The reflection $s_a\rightarrow -s_a$ corresponds to 
$x_a\rightarrow -x_a$ i.e. to $x \rightarrow x'$. }
\label{Figure2}
\end{center}
\end{figure}
\fi

\

For further use, note that the singular points of the moduli 
space (given by $s_a=0$) correspond to central holonomies, 
i.e. to holonomy operators 
$S_a=\pm 1$. 
There are $2^4=16$ such possibilities, as expected. If $S_a=\epsilon_a 1_2$, 
with $\epsilon_a\in \{-1,1\}$ is a central holonomy, then a a constant 
representative of the associated $V$-gauge equivalence class is given 
by ${\cal A}_a=\frac{1}{2}(1-\epsilon_a)\pi \sigma_2$, with 
$\sigma_2=\left(\begin{array}{cc}
0 & i\\
-i & 0
\end{array}\right)$ the second Pauli matrix.

\section{The moment map and Fayet-Iliopoulos terms}

We obtained a supersymmetric field theories with 8 supercharges and
$SU(2)_R$ R-symmetry, and may expect a triplet of Fayet-Iliopulos terms 
for each central generator of the (projected) gauge group. Naively, 
there appears to be no such central generator since the theory is intrinsically
nonabelian even in case of single D-particle on $T^4/\Z_2$. However, 
since the abstract formulation of our system is formally nothing but 
a supersymmetric quantum mechanics, although infinite-dimensional,
one may try to find a central generator at that level. In this 
section, we follow this idea for the case of $T^4/\Z_2$ orbifolds.
As we will discover, the particular nature of the $\Z_2$ projection 
dictates that the center of the projected gauge group is nontrivial
but is generated by {\em localized} gauge transformations in the D4-brane
basis. Thus, one is lead to consider Fayet-Ilioupoulos terms 
which -- in the D4-brane basis -- are localized
\footnote{A similar situation was considered in \cite{KS} in the 
case of tori of real dimension 1 and 2 , but without orbifolding. 
In that case, the  problem admits a series 
of drastic mathematical simplifications, in particular due to the fact that 
the singularities involved are rather mild. We thank A. Kapustin for bringing 
this reference to our attention.} at the 
fixed points of 
the $\Z_2$ action on $T'$ (In the D-particle picture, such terms are 
completely delocalized on the lattice $\Lambda$. See the appendix for a 
detailed description of the Fayet-Iliopoulos terms in the D-particle picture). 
The presence  of localized Fayet-Iliopoulos parameters in our situation 
appears to be the open-string analogue of the well-known phenomenon \cite{DHVW} of 
localization of blow-up modes in the conformal field theory of closed strings 
on orbifolds.

\subsection{Formal considerations}

The formal aspects of the situation are most clearly discussed at the 
level of the representation-independent formulation. For 
this purpose, we treat ${\cal X}^a$ as formal algebraic quantities. 

\

\noindent{\bf The moment map}

\

The projected symmetry group ${\cal G}$ 
acts by unitary symplectic automorphisms of the 
space of operators $X$, which has a natural quaternionic structure, induced 
from that of the vector space $V$ (since $X$ are vector-valued 
with vector indices corresponding to $V$). Therefore, 
we have an associated quaternionic moment map ${\vec m}$, taking 
$X$ into the Lie algebra $\Lie({\cal G})$ of ${\cal G}$.

To formulate this precisely, we choose $(e_a)_{a=1..4}$ 
to be the real basis associated to a complex orthonormal 
basis $\epsilon_k(k=1,2)$ of $V$:
\bea
\label{basis}
e_i=\epsilon_i \nn\\
e_{i+2}=i\epsilon_i \nn
\eea
($i=1,2$). Then $(e_a)_{a=1..4}$ is automatically orthonormal. 
The expression of the moment map is most transparent 
in the associated complex coordinates:
\bea
X^k={\cal X}^{k}+i{\cal X}^{k+2}\nn\\
X^{\overline k}={\cal X}^k-i{\cal X}^{k+2}~~\nn
\eea
($k=1,2$), where the complex and real parts of the moment map are:
\bea
m_c=[X^1,X^2]\nn\\
m_r=\frac{i}{2}([X^1,X^{\overline 1}]+[X^2,X^{\overline 2}])~~. \nn
\eea
Writing $m_c=m_1+im_2$ and $m_r=-im_3$, with $m_k$ hermitian matrices gives 
the usual real components $(m_1,m_2,m_3)={\vec m}$. 

Lowering indices with $g_{ij}=g_{{\overline i},{\overline j}}=0, 
g_{i,{\overline j}}=g_{{\overline i},j}=\frac{1}{2}\delta_{ij}$, we have
$X_i=\frac{1}{2}X^{\overline i}$, $X_{\overline i}=\frac{1}{2}X^i$, so that:
\bea
X_k:=\frac{1}{2}({\cal X}_{k}-i{\cal X}_{k+2})\nn\\
X_{\overline k}:=\frac{1}{2}({\cal X}_{k}+i{\cal X}_{k+2})~~\nn
\eea
($k=1,2$). 
The hermicity conditions are equivalent with 
$X_k^+=X_{\overline k}$. 
In terms of the covariant quantities $X_k, X_{\overline k}$, 
the real components of the moment map are 
$m_1=2\mu_1, m_2=2\mu_2, m_3=2\mu_3$
\footnote{We use the quantities $\mu_1,\mu_2,\mu_3$ in order to avoid some 
cumbersome factors later.}, with:
\bea
\mu_1=[X_2,X_1]+[X_{\overline 1},X_{\overline 2}] ~\nn\\
\mu_2=i([X_2,X_1]-[X_{\overline 1},X_{\overline 2}])~ \nn \\
\mu_3=[X_1,X_{\overline 1}]+[X_2,X_{\overline 2}] ~\nn ~~.
\eea
Defining $Q_{ab}:=[{\cal X}_a,{\cal X}_b] ~(a,b=1..4)$, we can rewrite this 
as:
\bea
\label{moment_real}
\mu_1=iQ^{(a)}_{14}\\
\mu_2=-iQ^{(a)}_{12} \nn\\
\mu_3=iQ^{(a)}_{13}~~, \nn
\eea
where $Q^{(a)}_{ab}:=\frac{1}{2}(Q_{ab}-{\tilde Q}_{ab})$ is the 
antiself-dual 
part of the antisymmetric tensor $Q_{ab}$. Here ${\tilde Q}_{ab}=\frac{1}{2}
\epsilon_{abcd}Q_{cd}$, with $\epsilon_{1234}=+1$.

\

\noindent{\bf The superpotential and Fayet-Iliopoulos terms}

\

In the basis (\ref{basis}) the scalar potential becomes:
\be
\label{potential_1'}
W=\frac{1}{\rho}Tr\sum_{a,b=1..4}{[{\cal X}_a,{\cal X}_b]
[{\cal X}_a,{\cal X}_b]}~~,
\ee
up to a prefactor which we ignore.
Let $\frac{1}{\sqrt{2}}\lambda_m$ be self-adjoint elements which 
form a (real) orthonormal basis of the Lie algebra of the 
gauge group ${\cal G}$ 
(and thus a complex orthonormal basis of its complexification), with 
respect to the scalar product $<A,B>=Tr(A^+B)$
\footnote{Below we will allow $m$ to take on continuous values as well. 
In this case the sums over $m$ should be interpreted as integrals and 
$\delta_{m,n}$ as meaning $\delta(m-n)$}.  
Then a direct computation as in \cite{JM} gives the identity:
\bdm
W=-\frac{8}{\rho}Tr({\vec \mu}^2)=
-\frac{4}{\rho}\sum_m{[Tr(\lambda_m{\vec \mu})]^2}~~.
\edm

Now let $\frac{1}{\sqrt{2}}\lambda_s$ be an orthonormal basis of the center 
$\Lie_0({\cal G})$ 
of the Lie algebra of ${\cal G}$, and $\lambda_n$ such that 
$\frac{1}{\sqrt{2}}\lambda_n,\frac{1}{\sqrt{2}}\lambda_s$ form an orthonormal 
basis of $\Lie({\cal G})$. Then one can introduce a triplet of auxiliary 
fields
${\vec D}_s$ for each $s$. After integrating out the auxiliary fields, one 
obtains the modified scalar potential:
\be
W'=-\frac{4}{\rho}
(\sum_{n}{[Tr(\lambda_n{\vec \mu})]^2} +\sum_{s}{[Tr(\lambda_s{\vec \mu})-
\vec{\xi}_s]^2}) 
\ee
where ${\vec \xi}_s$ are the associated triplets of Fayet-Iliopoulos 
parameters. Then the vacuum constraint $W'=0$ requires:
\bea
Tr(\lambda_n{\vec \mu})=0 ~\nn\\
Tr(\lambda_s{\vec \mu})=\vec{\xi}_s \nn 
\eea
i.e.:
\bdm
{\vec \mu}(X)=\frac{1}{2}\sum_s{\vec \xi}_s\lambda_s~~.
\edm

\subsection{The D4-brane representation}

\

\noindent{\bf The moment map}

\

After implementing the translation projection, ${\cal X}_a$ become covariant 
derivatives:
\bdm
{\cal X}_a=i{\cal D}_a=i(\partial_a-i{\cal A}_a),
\edm
while the commutators are:
\bdm
Q_{ab}=[{\cal X}_a,{\cal X}_b]=+i{\cal F}_{ab}
\edm
with:
\bdm
{\cal F}_{ab}=+i[{\cal D}_a,{\cal D}_b]=
\partial_a{\cal A}_b-\partial_b{\cal A}_a-i[{\cal A}_a,{\cal A}_b]~~.
\edm
Therefore, ${\vec \mu}$ becomes the multiplication operator by the 
function ${\vec \mu}(x)$ whose real components are (cf \ref{moment_real}):
\bea
\mu_1=-{\cal F}^{(a)}_{14}\nn\\
\mu_2={\cal F}^{(a)}_{12}\nn\\
\mu_3=-{\cal F}^{(a)}_{13}~~. \nn 
\eea
Here ${\cal F}^{(a)}$ is the antiself-dual part of 
${\cal F}$.

\

\noindent{\bf The central part of the gauge group in the case of $T^4/\Z_2$}

\

Let us determine  the central part 
$\Lie_0({\cal G})$ of the Lie algebra $\Lie({\cal G})$. It is immediate that 
there are two types of central elements of the Lie algebra, 
namely:
\be
\label{generators1}
\lambda(x)=f(x)1_2
\ee
with $f(-x)=f(x)$, and:
\be
\label{generators2}
\lambda_l(x)=\sigma_3\delta_{T'}(x-x_l)
\ee
where $x_l$ are the 16 fixed points of the $\Z_2$ action. 
(We normalize $\lambda_l$ 
such that $tr\int_{T}\lambda_l(x)\lambda_{l'}(x)=2\delta_{T'}(x_l-x_l')$). 

To justify this, let $\phi(x)$ be an element of $\Lie_0({\cal G})$ (this 
is an $u(2)$ matrix-valued function on $T'$). 
Since $\phi$ belongs to $\Lie({\cal G})$, it must satisfy the infinitesimal 
form of the projection condition on the gauge transformations:
\bdm
\sigma_3 \phi(x)\sigma_3=\phi(-x)~~.
\edm
Thus $\phi$ must have the form:
\bdm
\phi(x)=\left(\begin{array}{cc}
	e_1(x) & o_1(x)\\
	o_2(x) & e_2(x) 
	\end{array}\right)~~, \\
\edm
with $e_k,o_k$ even, respectively odd functions on $T'$. 

On the other hand, being a central element of $\Lie({\cal G})$, $\phi(x)$ 
must commute with all (projected) gauge transformations:
\be
\label{foobar}
U(x)\phi(x) U(x)^{-1}=\phi(x)~~.
\ee
Applying this condition for 
the constant $SU(2)$ gauge transformations $U(x)=e^{i\alpha\sigma_3}, 
~\alpha \in \R$ (which 
clearly satisfy the projection condition) and combining with the first 
relation shows that $\phi(-x)=\phi(x)$, which requires that $o_1(x)=o_2(x)=0$. 
Now consider a general $SU(2)$ gauge transformation satisfying the projection 
conditions:
\bdm
U(x)=\left(\begin{array}{cc}
	u(x) & v(x)\\
	-v^{*}(x) & u^{*}(x) 
	\end{array}\right)~~, \\
\edm
with $u(-x)=u(x), v(-x)=-v(x)$ and $|u(x)|^2+|v(x)|^2=1$. Then 
(\ref{foobar}) reduces to the constraint $e_1(x)v(x)=v(x)e_2(x)$. 
If $x$ is not a fixed point of the $\Z_2$ action, then $x \neq -x$ 
and it is always possible to construct a smooth {\em odd} 
function $v$ such that $|v|\leq 1$ and $v$ takes a nonzero value 
at $x$. Then we must have $e_1(x)=e_2(x):=f(x)$  and $U(x)=f(x)1_2$. 
However, if $x$ is a fixed point, then the evenness condition requires 
$v(-x)=-v(x)=v(x)$ (since $x=-x$), hence $v$ must be zero at $x$. 
It follows that  $\phi$ is constrained to be proportional to the identity 
matrix at all points except for the fixed points. At the fixed points 
$x_l (l=0..15)$, $\phi(x_l)$ can be any real diagonal matrix, 
i.e. any real linear combination of $1_2$ and $\sigma_3$. 
If one requires $\phi$ to be smooth over $T'$, this latter freedom is, of 
course, irrelevant. However, if one allows for singular $\phi(x)$, then 
one obtains the central generators 
(\ref{generators1},\ref{generators2}). 

\

\noindent{\bf The Fayet-Iliopoulos constraints}

\

For what follows we will set the Fayet-Iliopoulos
parameters associated to 
(\ref{generators1})  to zero, i.e. we impose the tracelessness constraint:
\bdm
tr({\vec \mu(x)})={\vec 0}~~.
\edm
This is automatically satisfied for $su(2)$ connections, which is the 
case we are considering. 
Then turning on Fayet-Iliopoulos parameters ${\vec \xi}_l$ 
associated to the 16 central generators (\ref{generators2}) leads to the 
vacuum constraints:
\be
\label{FI_constraint2}
\vec{\mu}(x)=\frac{1}{2}\sum_{l=0..15}{{\vec \xi}_l\sigma_3
\delta_{T'}(x-x_l)}~~,
\ee
i.e.:
\bea
\label{FIa}
{\cal F}^{(a)}_{14}=-\mu_1=-\frac{1}{2}\sum_l{\xi^1_l\sigma_3\delta_{T'}
(x-x_l)}~~~~\\
{\cal F}^{(a)}_{12}=\mu_2=\frac{1}{2}\sum_l{\xi^2_l\sigma_3\delta_{T'}
(x-x_l)}~\nn \\
{\cal F}^{(a)}_{13}=-\mu_3=-\frac{1}{2}\sum_l{\xi^3_l\sigma_3\delta_{T'}
(x-x_l)}~~ \nn .
\eea
The action of the symmetry group, the description of its center, and the 
form of the Fayet-Iliopoulos terms in the D-particle representation are 
discussed in the Appendix.

\section{The moduli space in the presence of Fayet-Iliopoulos terms}

\subsection{Naive considerations}

The equations ~(\ref{FIa}) completely determine $F^{(a)}$. 
Since we chose to work with
the regular representation on the Chan-Paton factors, the gauge
bundle in the D4-brane representation is trivial
(a fact that ensures, for example, that no additional topological
terms arise when interpreting the Hilbert space manipulations of 
subsection 4.1 in the gauge theory realization of subsection 4.2). 
In particular, this means
that the second Chern class of the gauge bundle vanishes, giving us 
a relation between  $F^{(s)}$ and $F^{(a)}$:
\bdm
\int_{T}{dx~[tr({\cal F}^{(s)}_{ ab }{\cal F}^{(s)}_{ab})-
tr({\cal F}^{(a)}_{ ab }{\cal F}^{(a)}_{ab})]}=0~~.
\edm
With our form of the Fayet-Iliopoulos constraints, the quantity 
$\int_{T}{dx~tr({\cal F}^{(a)}_{ ab }{\cal F}^{(a)}_{ab})}$ is divergent, 
so the same should be true of $\int_{T}{dx~tr({\cal F}^{(s)}_{ ab }
{\cal F}^{(s)}_{ab})}$. It seems reasonable to expect that the mechanism 
for obtaining such an infinity is common for $F^{(s)}$ and
$F^{(a)}$. Therefore, we propose that the self-dual part also 
has to be of the form for any vacuum configuration;
\bea
\label{FIs}
{\cal F}^{(s)}_{14}=\frac{1}{2}\sum_l{\eta^1_l\sigma_3\delta_{T'}(x-x_l)} 
~\nn\\
{\cal F}^{(s)}_{12}= \frac{1}{2}\sum_l{\eta^2_l\sigma_3\delta_{T'}(x-x_l)} 
~~ \\
{\cal F}^{(s)}_{13}=\frac{1}{2}\sum_l{\eta^3_l\sigma_3\delta_{T'}(x-x_l)}   
\nn ~.
\eea
If this is indeed the case, then cancellation of $tr({\cal F}^{(s)}_{ ab }
{\cal F}^{(s)}_{ab})$ and $tr({\cal F}^{(a)}_{ ab }{\cal F}^{(a)}_{ab})$ 
in $c_2$ occurs locally and requires that ${\vec \eta}_l$ satisfy:
\bdm
{\vec \eta}_l^2={\vec \xi}_l^2
\edm
for all $l=0..15$. That is, each ${\vec \eta}_l$ is constrained to lie on a 
$2$-sphere of radius $|\vec\xi_l|$. This gives 16 2-spheres, each of which
parametrizes local behaviour of $F$ at each of 16 fixed points, given the
FI parameters $\vec\xi_I$.

\vskip 5mm

To understand the situation better, note that, since the punctured 
torus $T'_p:=T'-\{x_0...x_{15}\}$ has 
$\pi_1=\Z^4$, a connection obeying ~(\ref{FIa}) has a well-defined 
parallel transport operator $P(v)$ as in section 3, 
which determines the connection uniquely. 
In particular, we have an associated holonomy representation 
$S:=P|_{\Lambda'}$. If a (complex) gauge in which ~(\ref{FIs}) holds exists, 
then, as in section 3, $S$ determines $A$ up to gauge transformations  
which are smooth over the {\em punctured} torus $T'_p$.
However, such a gauge transformation $U(x)$  may become  
singular at the puncture points $x_0..x_{15}$, so that $A$ is not 
determined by $S$ up to a gauge transformation which is smooth over 
{\em whole of} $T'$. Therefore, if -- as is natural -- 
we build the moduli space by using gauge transformations which are smooth 
over the entire $T'$, then the map $\psi$ from  gauge equivalence classes of  
connections (which are the points of the moduli space ${\cal M}_\xi$) 
to equivalence classes of holonomies modulo transformations of type 
(\ref{hol_gauge}) may not be one to one. In a gauge in which 
(\ref{FIa},\ref{FIs}) hold, fixing the equivalence class $[S]$ of the 
holonomy 
modulo transformations (\ref{hol_gauge}) determines $A$ only up to 
a gauge transformation $U(x)$ which may be singular at $x_0...x_{15}$.  
However, in order to fix $A$ up to a gauge transformation which is smooth 
over the entire $T'$, one also needs to specify the asymptotic behaviour of 
${\cal A}_a(x)$ as $x$ approaches any of the fixed points. This asymptotic 
behaviour is encoded by the parameters ${\vec \eta}_0..{\vec \eta}_{15}$. 

At this point, it seems reasonable that the moduli space 
${\cal M}_\xi$ can be identified with a subset of 
the direct product $(S^2)^{16}\times (T/\Z_2)$, where the first factor 
correponds to ${\vec \eta}_0..{\vec \eta}_{15}$, while the second factor 
corresponds to projected monodromies. Note, however, that ${\cal M}_\xi$
coincides with $(S^2)^{16}\times (T/\Z_2)$ only if the parameters 
${\vec \eta}_0..{\vec \eta}_{15}$ and $S$ can be varied independently. 
To show that ${\cal M}_\xi$ is at least 
toplogically equivalent to a K3 surface, 
it would suffice to show that the preimage $\psi^{-1}([S])$ contains exactly 
one point unless $[S]$ corresponds to a fixed point of $T/\Z_2$ (i.e. unless 
$S$ is a central holonomy), while $\psi^{-1}([S])$ would be a 2-sphere for 
each central $[S]$
To establish this, one may try to deduce a contraint between 
${\vec \eta}_0..{\vec \eta}_{15}$ and $[S]$, which would fix the values of  
${\vec \eta}_l$ for all noncentral $[S]$, while allowing for a nontrivial set 
of solutions ${\vec \eta}_0..{\vec \eta}_{15}$ 
(presumably parametrized by the points of a two-sphere) for a central $[S]$. 

Even granted that (\ref{FIa}) can be imposed 
(which is not clear by any means), 
implementing this idea is a rather difficult task, essentially due to the 
fact that the equations ~(\ref{FIa}) are not mathematically well-defined as 
they 
stand. Indeed, we obtained these equations by rather cavalier formal 
manipulations, but, in order to give them a clear 
meaning, one needs to specify a regularization of the commutators 
$[{\cal A}_a, {\cal A}_b]$ entering ${\cal F}_{ab}$. The reason is that, 
if one wants to satisfy (\ref{FIa}) in distributions, then $A(x)$ must 
scale at least as $\frac{1}{|x-x_l|^2}$ near the fixed points. In this case, 
the commutators $[{\cal A}_a, {\cal A}_b]$ scale at least as 
$\frac{1}{|x-x_l|^4}$, which means that they cannot be locally integrable 
functions around the fixed points 
(and thus do not define a distribution in a canonical way). 
Therefore, it is not immediately clear what the meaning of the object
$[{\cal A}_a, {\cal A}_b]$ is as a distribution. To make strict sense out of 
(\ref{FIa}), one needs to give an appropriate definition (`regularization' 
in the sense of distribution theory) of this commutator, 
compatible with the formal computations that we carried out in order to 
arrive at 
(\ref{FIa})\footnote{This can be viewed essentially as the problem of giving 
an adequate meaning to a certain type of distribution product.}. 

Alternatively, one can 
think about this in the D-particle representation as follows 
\footnote{The solution of the projection conditions as well as the 
counterparts of our Fayet-Iliopoulos terms in the D-particle respresentation 
are discussed in detail in the appendix.}. 
As discussed in the appendix, in order to solve the  Fayet-Iliopoulos
constraints ~(\ref{FIa}), one must include field 
configurations $A(t)$ which do not decrease fast at infinity. 
Specifying the precise 
class of such configurations is equivalent to specifying the class of 
connections ${\cal A}_a$ in which we look for a solution of ~(\ref{FIa}). 
Depending 
on the precise class of $A(t)$ chosen, one will obtain, via the Fourier 
transform, 
a different interpretation of ${\cal F}_{ab}$ as a distribution. 
We see that the above line of thinking leads to rather nontrivial technical 
problems. Therefore, in the 
next subsection we will give an alternate -- but still not rigorous -- 
argument.

\subsection{K3 surface}

One may hope that ${\cal M}_\xi$ would give a smooth and compact manifold 
which continuously reduces to the singular moduli space $T/\Z_2$ in the 
limit $\xi_l=0, \forall l=0\dots15$. Such a smooth and compact moduli space 
must be a K3 surface by a general argument. The equations ~(\ref{FIa}) 
present ${\cal M}_\xi$ as an  (infinite  form of a) hyperkahler quotient, 
where the space being quotiented is the affine space 
${\cal Q}$ of all (suitable) connections (the latter space has a natural 
hyperkahler structure, induced from the hyperkahler structure of $V$). 
Assuming that a suitable analogue of the argument of \cite{HKLR} 
could be made in our case, ${\cal M}_\xi$ inherits a hyperkahler structure 
from ${\cal Q}$, in a manner controlled by the FI parameters 
${\vec \xi}_0\dots{\vec \xi}_{15}$. Given that ${\cal M}_\xi$ continuously 
reduces to $T/\Z_2$ for $\xi_l \rightarrow 0$, it must be a a four-manifold. 
Since a smooth, compact and connected hyperkahler four-manifold can only be a 
complex two-dimensional torus or a K3 surface, and since ${\cal M}$ 
reduces to $T^4/\Z_2$ in the limit of zero Fayet-Iliopoulos terms, 
the resolved moduli space can only be a K3.

\subsection{Parameter count}

As a further check, let us see whether we have the correct number of moduli. 
The blow-up of $T/\Z_2$ at the 16 fixed points gives a Kummer surface $X$, 
carrying the famous 16 lines associated to the exceptional divisors
\cite{Barth}. These 
are rational curves $\Delta_l~(l=0..15)$ of self-intersection -2 and give 
elements of the Picard lattice. 
When talking about the blow-up, we are considering everything in the 
algebraic-geometric approach and therefore we fix a choice of complex
structure by taking a definite embedding in some projective space.
For a fixed complex structure, the Kahler form $\omega$  
is characterized by its periods $\Omega_l$ along the 
16 two-cycles $[\Delta_l]\in H_2(X,\R)$(the areas of the 2-spheres 
$\Delta_l$). This gives 16 real parameters characterizing the Kahler 
structure. To specify a {\em hyperkahler} 
structure on $X$, we have to give the periods of three Kahler forms 
$\omega^{(k)}(k=1..3)$ along $[\Delta_l]$. This gives a total of 48 real 
parameters, which is the same as the number of real parameters $\xi_l^{(k)}~
(l=0..15,k=1..3)$ controlling the moduli space ${\cal M}_\xi$.

\subsection{Sheaf interpretation}

The singular connections discussed above may appear to be less strange if one 
considers the problem from the point of view of holomorphic sheaves. For this, 
let us start by reformulating the moduli space of flat connections in  the 
holomorphic language. 
The Hitchin-Kobayashi correspondence (see, for example, \cite{DK}) 
tells us that giving a flat connection on the 
(differentiable) hermitian bundle $I$ is 
equivalent to giving a semistable holomorphic structure on $I$. 
This makes $I$ into a holomorphic vector bundle over $T'$. 
The tracelessness condition on the connection is equivalent to the 
requirement that the determinant line bundle $detI=\Lambda^2 I$ be 
holomorphically trivial.

Given any flat connection on $I$, one 
can find a gauge in which its connection matrix is constant. 
Since such a self-adjoint matrix is diagonalizable by a constant unitary gauge 
transformation, it follows that flat connections on $T'$ are reducible. 
Therefore, the associated holomorphic bundle $I$ decomposes as a direct sum 
of flat holomorphic line bundles:
\bdm
I=L\oplus L^{-1} ~~,
\edm
with $c_1(L)=0$. Thus, the moduli space of flat $SU(2)$ 
connections on $T'$ 
coincides with the moduli space of holomorphic line bundles $L$ on $T'$, which 
is isomorphic with the dual torus $T$. This is the holomorphic version of 
the the argument giving the moduli space of flat connections on $T'$. 
In the case of $T/\Z_2$, one has to consider the obvious 
equivariant adaptation of the above.  

It is a general fact that moduli spaces of self-dual connections over 
a compact manifold admit compactifications via moduli spaces of sheaves. 
The typical examples are Gieseker's compactification (via moduli spaces 
of Gieseker stable torsion-free sheaves) and the generally smaller 
compactification 
proposed by Maruyama\cite{Maruyama}, 
which is essentially obtained by restricting the 
class of sheaves allowed. Since torsion-free sheaves are essentially 
a singularization of vector bundles, one may expect that there exists 
a degenerate form of the Hitchin-Kobayashi correspondence, relating 
torsion-free sheaves to certain singular gauge connections.  
Indeed, such a correpondence was proposed in \cite{BS} for 
the case of reflexive sheaves (which are a particular class of torsion-free 
shaves). Moreover, it was noted in \cite{Maruyama} that the sheaves 
achieving the Maruyama compactification are related to singular limits of 
instantons. 

It is tempting, therefore, to suppose that the moduli space 
of connections obeying ~(\ref{FIa}) could be related to an appropriate 
class of holomorphic sheaves over $T'/\Z_2$. 
To make this slightly more explicit, consider setting $\xi^c_l$ to zero. 
If an appropriate version of the usual relation between Kahler and 
Geometric Invariant Theory quotients can be established in our case, 
then ${\cal M}_{(\xi_r,0)}$ 
could be identified with a moduli space of connections subject only 
to the (homogeneous) complex moment map equations, modulo {\em complexified} 
gauge transformations. 
Provided that a suitable sheaf interpretation of our connections can be 
found, this would coincide with the moduli space of an appropriate class 
of equivariant sheaves, similar to what happens in the case of moduli 
spaces of instantons over the noncompact orbifolds $\C^2/\Z_n$
\cite{CL1}. Since sheaf moduli spaces can 
be studied by the powerful methods of algebraic geometry, this may provide 
a better approach to the problem at hand. In this context, it may be noted 
that moduli spaces of sheaves often provide compact resolutions of moduli 
spaces of holomorphic bundles, and that there are situations 
(as that discussed in \cite{CL1}) when certain moduli spaces of 
equivariant sheaves realize the resolution of orbifolds. 
We hope to pursue this approach further elsewhere.

\section{Conclusions}

We gave a clear formulation of the effective field theory of one D-particle 
over $T^4/\Z_n$ orbifolds and an explicit derivation of the associated 
singular moduli space, as a moduli space of equivariant flat connnections 
in a product $V$-bundle over the orbifold.  
We showed that a straightforward modification of the procedure of 
\cite{DM,JM} leads to localized Fayet-Iliopoulos
 terms and to a theory whose vacua are 
described by an `infinite' version of a hyperkahler quotient.
We also presented some evidence that the resulting moduli 
space may provide a resolution of the compact orbifold.

\bigbreak\bigskip\bigskip\centerline{{\bf Aknowledgements}}\nobreak
\bigskip

We would like to  thank Philip Argyres, Robert Friedman, Igor Krichever, 
John Morgan, David Morrison, and Henry Tye for a number of very useful 
discussions. P.Y. is grateful to the Korea Institute for Advanced Studies
for hospitality.
The work of BRG is supported by DOE grant DE-FG02-92ER40699 and a National
Young Investigator award.
The work of CIL is supported by a Columbia University
Fister Fellowship and by the DOE grant DE-FG02-92ER40699B. 
The work of PY is supported by the National Science Foundation.

\

\

\appendix

\section{The form of the Fayet-Iliopoulos terms in the intermediate  
representation}

In the `intermediate' representation (\ref{intermediate}), the projection 
constraints on $X_{\alpha, \beta}(s,t):=(s,\alpha|X|t, \beta)\in V$ 
become:
\be
\label{Xproj_int}
\rho_{reg}(\alpha')X(\gamma(\alpha')s+t', \gamma(\alpha')t+t')=
\gamma(\alpha')X+2\pi t'1_n \delta_{s,t}~~,
\ee
where we combined the vectors $X_{\alpha, \beta}(s,t)$ into the $n$ by $n$ 
matrix $X(s,t)=(X_{\alpha,\beta}(s,t))_{\alpha,\beta=1..n}$. Defining 
$A(s):=X(s,0)$ as in \cite{Taylor}, the projection conditions by 
$\Z^4$ translations (obtained by taking $\alpha'=0$ in (\ref{Xproj_int})) 
require:
\be
\label{boo}
X(s,t)=A(s-t)+2 \pi t 1_n\delta_{s,t}~~,
\ee
while the $\Z_n$ projection (corresponding to $t'=0$) 
constrains $A(t)$ to satisfy:
\bdm
\rho_{reg}(\alpha)A(t)\rho_{reg}(-\alpha)=\gamma(\alpha)A(\gamma(-\alpha)t)~~. 
\edm
Equation (\ref{boo}) shows that $A_{\alpha,\beta}(s-t)=(s\alpha|A|t\beta)$, 
where $A$ is the Hilbert space operator used in section 2. 

Defining $U(s,t)=(U_{\alpha,\beta}(s,t))_{\alpha,\beta=1..n}$, 
with $U_{\alpha,\beta}(s,t)=(s,\alpha|U|t,\beta)$, the projection condition 
on the gauge group is:
\bdm
\rho_{reg}(\alpha')U(\gamma(\alpha')s+t', \gamma(\alpha')t+t')
\rho_{reg}(-\alpha')=U(s,t)~~. 
\edm
For $\alpha'=0$, this gives $U(s,t)=U(s-t)$, where $U(t):=U(t,0)$, while 
for $t'=0$ it becomes:
\bdm
\rho_{reg}(\alpha)U(t)\rho_{reg}(-\alpha)=U(\gamma(-\alpha)t)~~. 
\edm
In general, an operator $R$ satifies the projection condition by translations:
\be
\label{proj_by_trsl}
U(t,0)^{-1}RU(t,0)=R
\ee
if and only if its `intermediate' representation $R(s,t)=
((s,\alpha|R|t,\beta))_{\alpha,\beta=1..n}$ has the form:
\bdm
R(s,t)=R(s-t,0):=R(s-t)~~.
\edm
If $R,~S$ are two such operators, then their operator product $C=RS$ also 
satisfies (\ref{proj_by_trsl}) and its intermediate representation 
$C(s)=C(s,0)$ 
is given by:
\bdm
C=R*S~,
\edm
where $R*S$ is the convolution product of matrix valued  
sequences defined over $\Lambda$:
\bdm
(R*S)(t)=\sum_{u,v \in \Lambda, ~u+v=t}{R(u)S(v)}~~. 
\edm
In particular, unitarity of a gauge transformation $U$ is equivalent to the 
constraint: 
\bdm
U*U^+(t)=1_2, ~\forall t \in \Lambda~~,
\edm
(where $U^+(t)=U(-t)^+$), 
while its action on 
$A$ is reflected by the transformation:
\bdm
A(t) \rightarrow (U*A*U^{-1})(t)+2\pi \sum_{u+v=t}{U(u)vU^+(v)} ~~
\edm
of the matrix-valued sequence $A(t)$. Since the identifications used to 
build the moduli space are given by such nonlocal transformations, 
identifying the center of the projected gauge group in the 
D-particle picture is 
difficult. This is why we presented most of our discussion in the D4-brane 
description, where the relevant arguments are more transparent. Having 
identified the center of the gauge group ${\cal G}$ in that picture, one 
can wonder what it corresponds to in the more intuitive language of 
D-particles. One would also like to have an understanding in this language of 
the reason for the existence of the nontrivial central elements 
(\ref{generators2}). 

To answer these questions, let us first construct the 
D-particle analogues of our central generators. 
If $R$ is any operator obeying (\ref{proj_by_trsl}), then the change of basis 
(\ref{fourier_abstract}) between the intermediate and the D4-brane 
representations corresponds to the Fourier transform:
\bea
R(t)=\int_{T'}[dx]R(x)e^{2\pi i(t,x)}\nn\\
R(x)=\sum_{t \in \Lambda}{R(t)e^{-2\pi i(t,x)}}~~.\nn
\eea
(In the D4-brane representation, the projection condition by translations 
is equivalent to the requirement that $R$ be diagonal in the dual torus 
variables, i.e. 
$<x,\alpha|R|y,\beta>=R_{\alpha,\beta}(x)\delta_{T'}(x-y)$. Then $R(x)$ in 
the above relation is defined by 
$R(x)=(R_{\alpha,\beta}(x))_{\alpha,\beta=1..n}$.)

Now consider the case $n=2$. 
Then the $\Z_2$ projection on the gauge group is (in the intermediate 
representation):
\bdm
\sigma_3 U(t)\sigma_3=U(-t)~~.
\edm 
In the D4-brane representation, an element of the Lie algebra of 
${\cal G}$ is a matrix-valued function $\lambda(x)$ on $T'$, associated 
to an operator $\lambda$ in ${\cal H}$ which obeys (\ref{proj_by_trsl}). 
Hence in the intermediate representation it corresponds to a matrix-valued 
sequence:
\bdm
\lambda(t)=\int_{T'}[dx]\lambda(x)e^{2\pi i(t,x)}~~.
\edm
Therefore, the diagonal generators (\ref{generators1}) give sequences 
of the form:
\be
\label{generators1_intmd}
\lambda(t)={\hat f}(t)1_2~~,
\ee
with ${\hat f}(t)=\int_{T'}[dx]f(x)e^{2\pi i(t,x)}$ the Fourier transform of 
the scalar function $f$, while the other 16 generators correspond to:
\be
\label{generators2_intmd}
\lambda_l(t)=\sigma_3e^{2\pi i (t,x_l)}~~.
\ee
To understand directly why the latter generators are central in the 
projected gauge group, note that any fixed point $x_l \in T'$ is induced 
from a point $m_l\in \frac{1}{2}\Lambda'$ on the covering space ($m_l$
is determined up to an element of $\Lambda'$). Therefore, we have 
$(t,x_l)\equiv (t,m_l) ({\rm ~mod~} \Z) \in \frac{1}{2}\Z, 
~\forall t \in \Lambda$, 
so that $ e^{2\pi i (t,x_l)}=e^{-2\pi i (t,x_l)}$. Moreover, we have 
$(2t, x_l)\equiv (2t,m_l) ({\rm ~mod~} \Z) \in \Z,~\forall t \in \Lambda$, 
which implies 
$ e^{2\pi i (v,x_l)}= e^{2\pi i (v-2t,x_l)},~\forall v \in \Lambda$. 
Using these two facts, we can see directly that $\lambda_l$ commutes with 
any element $U$ of the projected gauge group:
\begin{eqnarray*}
(U*\lambda_l)(t)=\sum_{u+v=t}{U(u)\sigma_3 e^{2\pi i (v,x_l)}}=  \\
\sum_{u+v=t}{\sigma_3 U(-u)e^{2\pi i (v,x_l)}}=
\sum_{u+v=-t}{\sigma_3 e^{2\pi i (v,x_l)}U(u)}= \\ 
\sum_{u+v=t}{\sigma_3e^{2\pi i (v,x_l)}U(u)}=(\lambda_l*U)(t)~~.
\end{eqnarray*}
In the second line we made the change of variables 
$u\rightarrow -u, v\rightarrow -v$ and used the first property discussed 
above, while in the third line we made the change of variables 
$v \rightarrow v -2t$ and used the second property. 

If $l=0$, then $x_0=0$ and the sequence $\lambda_0$ is constant:
\bdm
\lambda_0(t)=\sigma_3, ~\forall t~~.
\edm
If $l=1..15$, then $m_l$ must belong to $\frac{1}{2}\Lambda'-\Lambda'$ and 
the sequence  $\lambda_l$ is never constant. It consists of alternating 
matrices $+\sigma_3, -\sigma_3$:
\bdm
\lambda_l=(-1)^{2(t,x_l)}\sigma_3~~.
\edm
To make this more explicit, choose dual integral bases $(\tau_1..\tau_4)$, 
$(\tau'_1..\tau'_4)$ of $\Lambda, \Lambda'$:
\bdm
(\tau_a,\tau'_b)=\delta_{ab},~\forall a,b=1..4~~.  
\edm
Pick representatives $m_r:=\frac{1}{2}(r_1\tau'_1+...+r_4\tau'_4)$ of 
the fixed points, with $r=(r_1...r_4)$, $r_a\in \{0,1\}\approx \Z_2,
~\forall a=1..4$. 
This corresponds to indexing the fixed points by $r\in (\Z_2)^4$.  
Then:
\bdm
\lambda_r(t)=\sigma_3(-1)^{r_1t^1+...r_4t^4}~~,
\edm
for all $t=t^1\tau_1+..+t^4\tau_4 \in \Lambda$. A two-dimensional model of 
the situation is given below. 

\

\iffigs
\begin{figure}[htpb]
\begin{center}
\input{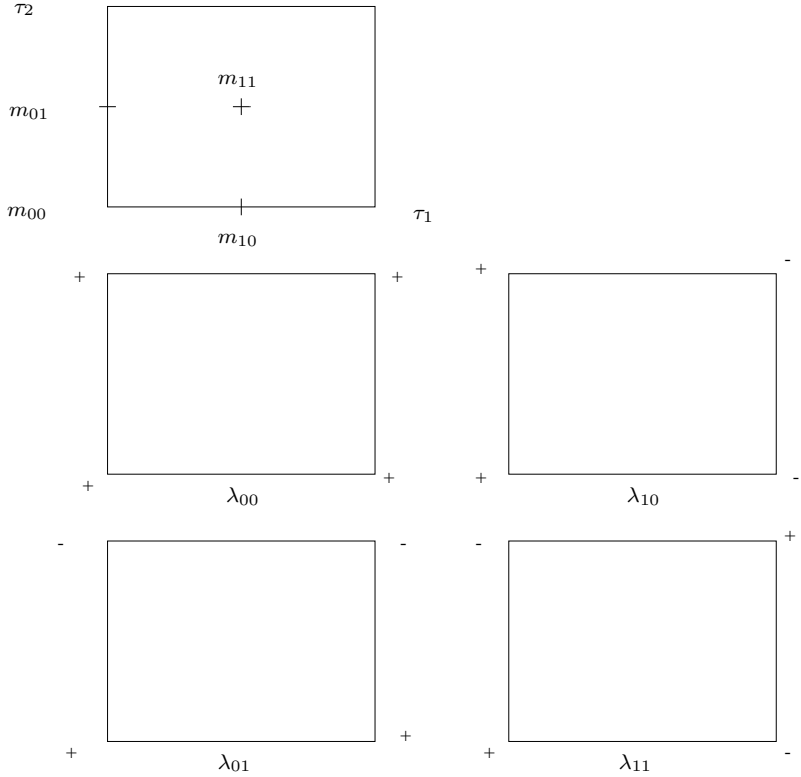}
\caption{\footnotesize The sign prefactors of the 
first components of $\lambda_r$.}
\label{Figure3}
\end{center}
\end{figure}
\fi

\

As expected, the Fourier transform maps the localized central elements 
(\ref{generators2}) in the D4-brane picture into completely delocalized 
elements (\ref{generators2_intmd}). The singularities in the gauge connection 
needed to obey the Fayet-Iliopoulos constraints (\ref{FI_constraint2}) in 
the D4-brane picture are 
reflected in the necessity to include delocalized configurations $A(t)$, i.e. 
configurations for which $A(t)$ does not decrease fast enough at infinity. 
The fact that the  Fayet-Iliopoulos terms are delocalized in the D-particle 
picture means 
that they are essentially a `large volume' effect. The ultimate reason for 
this is, of course, the fact that we projected the system by the integral
translation group $\Z^4$. Once this projection is implemented (at least in 
the regular representation, as we did in this paper), the only 
localized central generators still allowed in the D-particle picture are the  
diagonal elements (\ref{generators1_intmd}). 

Finally, let us write down the Fayet-Iliopoulos constraints in the 
intermediate representation. The abstract operators 
$Q_{ab}=[{\cal X}_a,{\cal X}_b]$ of section 4 
obey (\ref{proj_by_trsl}) and are given by:
\bdm
Q_{ab}(t):=Q_{ab}(t,0)=[{\cal A}_a,{\cal A}_b]_*(t)+2 \pi(t_a {\cal A}_b-
t_b {\cal A}_a)~~,
\edm
where 
$[{\cal A}_a,{\cal A}_b]_*:={\cal A}_a*{\cal A}_b -{\cal A}_b*{\cal A}_a$. 
Hence ${\vec \mu}$ also obeys (\ref{proj_by_trsl}) and gives a sequence 
${\vec \mu}(t):={\vec \mu}(t,0)$, with:
\bea
\mu_1(t)=iQ^{(a)}_{14}(t)\nn\\
\mu_2(t)=-iQ^{(a)}_{12}(t)\nn\\
\mu_3(t)=iQ^{(a)}_{13}(t)~~.\nn
\eea
Then the moment map constraints are given by:
\bdm
{\vec \mu}(t)=\frac{1}{2}\sum_{r \in (\Z_2)^4}{\vec \xi}_r \lambda_r(t)~~.
\edm
This is a system of nonlocal, nonlinear equations for ${\cal A}_a(t)$.

\end{document}